\input harvmac
\def\journal#1&#2(#3){\unskip, \sl #1\ \bf #2 \rm(19#3) }
\def\andjournal#1&#2(#3){\sl #1~\bf #2 \rm (19#3) }

\def\ie{{\it i.e.}}
\def\eg{{\it e.g.}}

\def\frac#1#2{{#1\over#2}}

\def\half{\frac12}

\def\inbar{\,\vrule height1.5ex width.4pt depth0pt}
\def\IC{\relax\hbox{$\inbar\kern-.3em{\rm C}$}}
\def\IR{\relax{\rm I\kern-.18em R}}
\def\IP{\relax{\rm I\kern-.18em P}}
\def\IZ{\relax{\rm I\kern-.18em Z}}

%
%

%
\catcode`\@=11
\def\slash#1{\mathord{\mathpalette\c@ncel{#1}}}
\overfullrule=0pt

\def\DD{{\cal D}}

\def\II{{\cal I}}
\def\JJ{{\cal J}}

\def\MM{{\cal M}}

\def\RR{{\cal R}}
\def\SS{{\cal S}}

\def\XX{{\cal X}}

\def\underrel#1\over#2{\mathrel{\mathop{\kern\z@#1}\limits_{#2}}}

\catcode`\@=12


%

\def\exp{{\rm exp}}


\def\twoone{{(2,1)}}

\rightline{RI-5-01, EFI-01-19}
\Title{
\rightline{hep-th/0106004}}
{\vbox{\centerline{Notes on $AdS_3$}}}
\medskip
\centerline{\it Amit Giveon${}^{1}$ and David Kutasov${}^{2,3}$}
\bigskip
\centerline{${}^1$Racah Institute of Physics, The Hebrew University}
\centerline{Jerusalem 91904, Israel}
\centerline{giveon@vms.huji.ac.il}
\smallskip
\centerline{${}^2$Department of Physics, University of Chicago}
\centerline{5640 S. Ellis Av., Chicago, IL 60637, USA }
\centerline{kutasov@theory.uchicago.edu}
\smallskip
\centerline{${}^3$Department of Physics, Weizmann Institute of Science}
\centerline{Rehovot 76100, Israel}

\bigskip\bigskip\bigskip
\noindent
We use the conjectured strong-weak coupling worldsheet duality
between the $SL(2)/U(1)$ and Sine-Liouville conformal field
theories to study some properties of degenerate operators and to
compute correlation functions in CFT on $AdS_3$. The same quantities
have been computed in the past by other means. The agreement between
the different approaches provides new evidence for the duality.
We also discuss the supersymmetric analog of this duality, the
correspondence between SCFT on the cigar and $N=2$ Liouville.
We show that in the spacetime CFT dual to string theory on $AdS_3$
via the AdS/CFT correspondence, the central term in the Virasoro
algebra takes different values in different sectors of the theory.
In a companion paper we use the results described here to study
D-branes in $AdS_3$.

\vfill

\Date{5/01}

\newsec{Introduction}

\lref\gks{A.~Giveon, D.~Kutasov and N.~Seiberg,
``Comments on string theory on AdS(3),''
Adv.\ Theor.\ Math.\ Phys.\  {\bf 2}, 733 (1998)
[hep-th/9806194].}
\lref\mms{J.~Maldacena, G.~Moore and A.~Strominger,
``Counting BPS black holes in toroidal type II string theory,''
hep-th/9903163.}
\lref\ks{D.~Kutasov and N.~Seiberg,
``More comments on string theory on AdS(3),''
JHEP {\bf 9904}, 008 (1999)
[hep-th/9903219].}
\lref\sw{N.~Seiberg and E.~Witten,
``The D1/D5 system and singular CFT,''
JHEP {\bf 9904}, 017 (1999)
[hep-th/9903224].}
\lref\mo{J.~Maldacena and H.~Ooguri,
``Strings in AdS(3) and SL(2,R) WZW model. I,''
hep-th/0001053.}
\lref\ags{R.~Argurio, A.~Giveon and A.~Shomer,
``Superstrings on AdS(3) and symmetric products,''
JHEP {\bf 0012}, 003 (2000)
[hep-th/0009242].}%
\lref\tesch{J.~Teschner,
``On structure constants and fusion rules in the SL(2,C)/SU(2) WZNW model,''
Nucl.\ Phys.\ B {\bf 546}, 390 (1999)
[hep-th/9712256].}
\lref\gksc{A. Giveon, D. Kutasov and A. Schwimmer,
``Comments on D-branes in $AdS_3$,'' [hep-th/0106005].}
\lref\nw{C.~R.~Nappi and E.~Witten,
``A Closed, expanding universe in string theory,''
Phys.\ Lett.\ B {\bf 293}, 309 (1992)
[hep-th/9206078].}
\lref\cosmo{Cosmo}
\lref\efr{S.~Elitzur, A.~Forge and E.~Rabinovici,
``Some global aspects of string compactifications,''
Nucl.\ Phys.\ B {\bf 359}, 581 (1991).}
\lref\msw{G.~Mandal, A.~M.~Sengupta and S.~R.~Wadia,
``Classical solutions of two-dimensional string theory,''
Mod.\ Phys.\ Lett.\ A {\bf 6}, 1685 (1991).}
\lref\witten{E.~Witten,
``On string theory and black holes,''
Phys.\ Rev.\ D {\bf 44}, 314 (1991).}
\lref\gkp{A.~Giveon, D.~Kutasov and O.~Pelc,
``Holography for non-critical superstrings,''
JHEP {\bf 9910}, 035 (1999)
[hep-th/9907178].}
\lref\dvv{R.~Dijkgraaf, H.~Verlinde and E.~Verlinde,
``String propagation in a black hole geometry,''
Nucl.\ Phys.\ B {\bf 371}, 269 (1992).}
\lref\abks{O.~Aharony, M.~Berkooz, D.~Kutasov and N.~Seiberg,
``Linear dilatons, NS5-branes and holography,''
JHEP {\bf 9810}, 004 (1998)
[hep-th/9808149].}
\lref\gk{A.~Giveon and D.~Kutasov,
``Little string theory in a double scaling limit,''
JHEP {\bf 9910}, 034 (1999)
[hep-th/9909110].}
\lref\gktwo{A.~Giveon and D.~Kutasov,
``Comments on double scaled little string theory,''
JHEP {\bf 0001}, 023 (2000)
[hep-th/9911039].}
\lref\egk{S.~Elitzur, A.~Giveon and D.~Kutasov,
``Branes and N = 1 duality in string theory,''
Phys.\ Lett.\ B {\bf 400}, 269 (1997)
[hep-th/9702014].}
\lref\egkrs{S.~Elitzur, A.~Giveon, D.~Kutasov, E.~Rabinovici and
G.~Sarkissian,
``D-branes in the background of NS fivebranes,''
JHEP {\bf 0008}, 046 (2000)
[hep-th/0005052].}%
\lref\sv{A.~Strominger and C.~Vafa,
``Microscopic Origin of the Bekenstein-Hawking Entropy,''
Phys.\ Lett.\ B {\bf 379}, 99 (1996)
[hep-th/9601029].}
\lref\strominger{A.~Strominger,
``Black hole entropy from near-horizon microstates,''
JHEP {\bf 9802}, 009 (1998)
[hep-th/9712251].}
\lref\fzzsl{V.~Fateev, A.~Zamolodchikov and A.~Zamolodchikov, unpublished.}
\lref\hk{K.~Hori and A.~Kapustin,
``Duality of the fermionic 2d black hole and N = 2 Liouville theory as
mirror symmetry,''
hep-th/0104202.}
\lref\kkk{V.~Kazakov, I.~K.~Kostov and D.~Kutasov,
``A matrix model for the two dimensional black hole,''
hep-th/0101011.}
\lref\adg{I.~Antoniadis, S.~Dimopoulos and A.~Giveon,
``Little string theory at a TeV,''
hep-th/0103033.}
\lref\gn{G. Giribet and C. Nunez, ``Correlators in $AdS_3$ string theory,''
hep-th/0105200.}
\lref\fh{T. Fukuda and K. Hosomichi, ``Three-point Functions
in Sine-Liouville Theory,'' hep-th/0105217.}
\lref\zaza{A.~Zamolodchikov and A.~Zamolodchikov,
``Liouville field theory on a pseudosphere,''
hep-th/0101152.}
\lref\kpz{V.~G.~Knizhnik, A.~M.~Polyakov and A.~B.~Zamolodchikov,
``Fractal Structure Of 2d-Quantum Gravity,''
Mod.\ Phys.\ Lett.\ A {\bf 3}, 819 (1988).}
\lref\gm{P.~Ginsparg and G.~Moore,
``Lectures On 2-D Gravity And 2-D String Theory,''
[hep-th/9304011].}
\lref\sahak{D.~Kutasov and D.~A.~Sahakyan,
``Comments on the thermodynamics of little string theory,''
JHEP {\bf 0102}, 021 (2001) [hep-th/0012258].}
\lref\MaldacenaUZ{
J.~Maldacena, J.~Michelson and A.~Strominger,
``Anti-de Sitter fragmentation,''
JHEP {\bf 9902}, 011 (1999)
[hep-th/9812073].
}
\lref\PetropoulosNC{
P.~M.~Petropoulos,
``String theory on AdS(3): Some open questions,''
hep-th/9908189.
}
\lref\ArgurioIR{
R.~Argurio, A.~Giveon and A.~Shomer,
``String theory on AdS(3) and symmetric products,''
Fortsch.\ Phys.\  {\bf 49}, 409 (2001)
[hep-th/0012117].
}

In this paper we study string theory on $AdS_3$ (the infinite cover
of the $SL(2)$ group manifold). This system received a lot of
attention over the years (see \eg\
\refs{\gks,\ks,\PetropoulosNC,\mo,\ArgurioIR} for some
recent discussions and additional references), for a variety of reasons.
Some of the reasons for our interest in this model are:
\item{1.}
We regard it as a warmup exercise for the study of string theory
in time dependent backgrounds. For instance, the coset
${[SL(2)\times SU(2)]/R^2}$ is a cosmological background
corresponding to a closed universe which begins and ends with
a singularity \nw. $AdS_3$ CFT is an important ingredient in
analyzing the physics of this model.
\item{2.}
It is relevant for the study of asymptotically linear dilaton theories,
such as the Liouville model and the cigar CFT \efr, which describes a
Euclidean two dimensional black hole \refs{\msw,\witten}. Liouville
can be obtained from $SL(2)$ (after a certain twist) by gauging a Borel
subgroup. The cigar corresponds to the coset $SL(2)/U(1)$ where the
$U(1)$ is associated with the compact timelike direction in $SL(2)$.
Gauging the non-compact spacelike $U(1)$ gives a Lorentzian two
dimensional black hole.
\item{3.}
Linear dilaton models have many applications. Liouville theory plays
an important role in two dimensional string theory, which is holographically
dual to a certain matrix quantum mechanics in the large $N$ limit
(see \eg\ \gm\ for a review). The cigar appears both in two dimensional
string theory, where it describes the high energy thermodynamics (see \kkk\
for a recent discussion) and in Little String Theory (LST), where it
again plays a role in the thermodynamics (see \eg\ \sahak),
and in a certain double scaling limit defined in \gk.
The near-horizon geometry of NS fivebranes in the presence of fundamental
strings interpolates between a linear dilaton region far from the strings
and an $AdS_3$ geometry near the strings \gkp.
D-branes stretched between non-parallel fivebranes that give rise at
low energies to four dimensional $N=1$ SYM \egk\ are described in the
near-horizon region of the intersection of the fivebranes as D-branes
living near the tip of the cigar \refs{\egkrs,\adg}.
One of the main motivations for this
work is to develop tools for studying such D-branes, with the hope of
learning more about $N=1$ SYM.
\item{4.} String theory on $AdS_3$ is an interesting special case of the
AdS/CFT correspondence. It does not require turning on RR backgrounds, and
thus can be studied (at weak string coupling)
by standard worldsheet techniques.
Also, the spacetime CFT is in this case two dimensional;
hence the corresponding
conformal symmetry is infinite dimensional. This symmetry can be realized
directly in string theory \refs{\gks,\ks} and its presence might provide
clues for the study of broken infinite dimensional symmetries in
string theory in general.
\item{5.}
String theory on $AdS_3$ is relevant for the study of the quantum mechanics
of $d=3,4,5$ black holes \refs{\sv,\strominger}.

\noindent
This is the first of two papers in which we discuss some aspects
of the dynamics of strings on $AdS_3$. The main issues that we
address here are the following:

\vskip .1in
\noindent
{\it Non-locality of string theory on $AdS_3$:}
the spacetime CFT corresponding to string theory on $AdS_3$ via the
AdS/CFT correspondence has some non-local features\foot{In
the absence of RR backgrounds. The system is not understood in
situations where RR backgrounds are turned on.}. Some of the
manifestations of this non-locality are:
\item{(a)} The spectrum of scaling dimensions in the spacetime CFT
contains a continuum above a finite gap. This continuum
corresponds to long strings living near the boundary of $AdS_3$
\refs{\gks,\MaldacenaUZ,\ks,\sw,\mo,\ags}. These strings do not
correspond to local operators in the spacetime CFT.
\item{(b)} Correlation functions in string theory on $AdS_3$ exhibit
singularities at values of the scaling dimensions where short strings
can scatter into long strings \refs{\gk,\gktwo}. Thus, the non-locality
associated with long strings influences the physics of short strings as well.
\item{(c)} As mentioned above, after gauging a $U(1)$, one finds
asymptotically linear dilaton theories such as Liouville and the cigar.
The latter are related to LST and are non-local in spacetime.

\noindent
In these notes we add another entry to this list. We show that
the central term of the spacetime Virasoro algebra, which was constructed in
\refs{\gks,\ks}, is given by a dimension zero operator which is not proportional
to the identity. The central charge is different in different sectors of
the theory; in a state with spacetime scaling dimension $h$, the central
charge has a contribution that grows like $h$. This generalizes observations
in \refs{\gks,\ks}, where it was shown that long strings carry non-zero
central charge; we will see that short strings carry central charge as well.

\vskip .1in
\noindent
{\it The conjectured duality between $SL(2)/U(1)$ and Sine-Liouville:}
V. Fateev, A. B. Zamolodchikov and Al. B. Zamolodchikov conjectured
\fzzsl\ that the CFT's on the cigar and Sine-Liouville theory are
equivalent under a strong-weak coupling duality. The detailed
statement of the duality appears in \kkk. In the supersymmetric case,
a similar duality between the supersymmetric $SL(2)/U(1)$ coset and
$N=2$ Liouville was proposed in \gk\ based on considerations involving
string dynamics near singularities of Calabi-Yau manifolds. Recently, there
was some more work on these dualities \refs{\hk,\gn,\fh}. We show here
that, assuming the duality and using general properties of degenerate
operators in CFT on $AdS_3$, leads to results for certain OPE coefficients
and correlation functions on $AdS_3$ which agree with those obtained by
solving the null state equations together with the Knizhnik-Zamolodchikov
equations for the current algebra blocks. This provides new evidence for
the duality.

\nref\gmoms{A.~Gerasimov, A.~Morozov, M.~Olshanetsky, A.~Marshakov and
S.~Shatashvili,
``Wess-Zumino-Witten Model As A Theory Of Free Fields,''
Int.\ J.\ Mod.\ Phys.\ A {\bf 5}, 2495 (1990).}

\nref\BernardIY{
D.~Bernard and G.~Felder,
``Fock Representations And Brst Cohomology In Sl(2) Current Algebra,''
Commun.\ Math.\ Phys.\  {\bf 127}, 145 (1990).
}
\nref\BershadskyIN{
M.~Bershadsky and D.~Kutasov,
``Comment on gauged WZW theory,''
Phys.\ Lett.\ B {\bf 266}, 345 (1991).
}
\nref\FurlanBY{
P.~Furlan, A.~C.~Ganchev, R.~Paunov and V.~B.~Petkova,
``Reduction of the rational spin sl(2,C) WZNW conformal theory,''
Phys.\ Lett.\ B {\bf 267}, 63 (1991).}
\nref\gawed{K.~Gawedzki,
``Noncompact WZW conformal field theories,''
hep-th/9110076.}
\nref\FurlanMM{
P.~Furlan, A.~C.~Ganchev, R.~Paunov and V.~B.~Petkova,
``Solutions of the Knizhnik-Zamolodchikov equation with rational isospins
and the reduction to the minimal models,''
Nucl.\ Phys.\ B {\bf 394}, 665 (1993)
[hep-th/9201080].
}
\nref\AwataSM{
H.~Awata and Y.~Yamada,
``Fusion rules for the fractional level sl(2) algebra,''
Mod.\ Phys.\ Lett.\ A {\bf 7}, 1185 (1992).
}
\nref\becker{K.~Becker and M.~Becker,
``Interactions in the SL(2,IR) / U(1) black hole background,''
Nucl.\ Phys.\ B {\bf 418}, 206 (1994)
[hep-th/9310046].}
\nref\AndreevBJ{
O.~Andreev,
``Operator algebra of the SL(2) conformal field theories,''
Phys.\ Lett.\ B {\bf 363}, 166 (1995)
[hep-th/9504082].
}
\nref\PetersenHS{
J.~L.~Petersen, J.~Rasmussen and M.~Yu,
``Fusion, crossing and monodromy in conformal field theory based on SL(2)
current algebra with fractional level,''
Nucl.\ Phys.\ B {\bf 481}, 577 (1996)
[hep-th/9607129].
}
\nref\FurlanVU{
P.~Furlan, A.~C.~Ganchev and V.~B.~Petkova,
``$A_1~{(1)}$ Admissible Representations -- Fusion Transformations and
Local Correlators,''
Nucl.\ Phys.\ B {\bf 491}, 635 (1997)
[hep-th/9608018].
}
\nref\tesss{J.~Teschner,
``Operator product expansion and factorization in the H-3+ WZNW model,''
Nucl.\ Phys.\ B {\bf 571}, 555 (2000)
[hep-th/9906215].}
\nref\ios{N.~Ishibashi, K.~Okuyama and Y.~Satoh,
``Path integral approach to string theory on AdS(3),''
Nucl.\ Phys.\ B {\bf 588}, 149 (2000)
[hep-th/0005152].}
\nref\hos{K.~Hosomichi, K.~Okuyama and Y.~Satoh,
``Free field appoach to string theory on AdS(3),''
Nucl.\ Phys.\ B {\bf 598}, 451 (2001)
[hep-th/0009107].}
\nref\kazhos{K. Hosomichi and Y. Satoh, ``Operator Product Expansion in String
Theory on $AdS_3$,'' hep-th/0105283.}

\vskip .1in
\noindent
{\it Properties of degenerate operators in CFT on $AdS_3$:}
We compute the current algebra blocks that enter four point
functions of two degenerate operators and two general ones. This
was already done\foot{For other work on correlation functions in
$SL(2)$ CFT see \eg\
\refs{\gmoms - \kazhos, \gn}.}
in \tesch, but we review the calculation here for two reasons.
One is that this is needed for comparing to the results
obtained using the duality mentioned in the previous paragraph.
In addition, these blocks are needed for studying D-branes on $AdS_3$,
which we do in \gksc.

\vskip .1in
These notes are organized as follows. In section 2 we begin with a very
brief review of CFT on $AdS_3$. We establish the notations and
quote some results that are needed later. In section 3 we discuss
the central charge of the spacetime Virasoro algebra, which corresponds
to the zero momentum dilaton. We show that
this operator is not proportional to the identity operator in string theory
on $AdS_3$ and compute its correlation functions with other operators.
We explain the interpretation of the results in terms of
the spacetime CFT and clarify the relation between the spacetime central
charge operator and the Wakimoto screening operator in the free field
realization of the model.

In section 4 we turn to a discussion of the properties of degenerate
operators in CFT on $AdS_3$. There is an infinite set of such operators,
labeled by two positive integers $(r,s)$. We consider in detail two such
operators, corresponding to $(r,s)=(1,2),\;(2,1)$, which can be thought
of as the generators of the set.
We show that the OPE's of these operators with
other primaries contain a finite number of terms, and compute the structure
constants by using the fact that they are dominated by the region near
the boundary of $AdS_3$, where one can use perturbation theory in either
the Wakimoto screening operator or the Sine-Liouville coupling. We also
show that one can use the resulting structure constants to compute correlation
functions in $SL(2)$ CFT and the results are in agreement with other methods
of computing them. This provides a test of the duality of \fzzsl.

In section 5 we compute the current algebra blocks corresponding to the
four point functions of two (identical) degenerate operators and two
(identical) generic ones. This calculation originally appeared in
\tesch, and is reviewed here for reasons that were mentioned above.

In section 6 we generalize the results
to the supersymmetric case, where our results provide evidence for the
duality of \gk. Some useful formulae are collected in an appendix.

\newsec{Some properties of CFT on $AdS_3$}

The WZNW model on $AdS_3$ is described by the following Lagrangian,
written in Poincar\'{e} coordinates $(u,\gamma,\bar\gamma)$
\eqn\lpoi{\CL=2k\left({1\over u^2}\partial u\bar\partial u
+u^2\bar\partial\gamma\partial\bar\gamma\right)~.}
The parameter $k$ is related to the radius of curvature of the space, $l$,
via\foot{We set the string length $l_s=1$.}  $k=l^2$. The boundary of $AdS_3$
is the two dimensional space labeled by $(\gamma,\bar\gamma)$ at $u\to\infty$.
In the Lorentzian case $\gamma$ and $\bar\gamma$ are independent real
coordinates. In the Euclidean case $\bar\gamma$ is the complex conjugate of
$\gamma$, and the boundary is the complex plane, or two-sphere.

The model described by the Lagrangian \lpoi\ is invariant under two copies
of the $SL(2,R)$ current algebra. The left moving symmetry is generated by
the currents $J^a(z)$, with $a=3,\pm$, satisfying the OPE algebra
\eqn\opealg{\eqalign{
J^3(z) J^\pm(w)\sim &{\pm J^\pm(w)\over z-w} \cr
J^3(z) J^3(w)\sim &-{{k\over2}\over (z-w)^2}\cr
J^-(z)J^+(w)\sim &{k\over (z-w)^2}+{2J^3(w)\over z-w}.\cr}}
A similar set of OPE's holds for the right moving $SL(2)$ current algebra.
The level $k$ of the current algebra \opealg\ is related to the central
charge of the CFT \lpoi\ via
\eqn\centch{c={3k\over k-2}.}
One is typically interested in $k>2$.

A natural set of observables is given by the eigenfunctions
of the Laplacian on $AdS_3$,
\eqn\limphin{\eqalign{
\Phi_h=&{1-2h\over\pi}\left({1 \over |\gamma-x|^2e^{Q\phi\over2}+
e^{-{Q\phi\over2}}}
\right)^{2h} = \cr
&-e^{Q(h-1)\phi} \delta^2(\gamma-x) + \CO(e^{Q(h-2)\phi}) +
{(1-2h)e^{-Qh\phi} \over \pi |\gamma-x|^{4h}} + \CO(e^{-Q(h+1) \phi}), }}
where\foot{Note that in \limphin\ we have rescaled $\phi$
and $\Phi_h$ relative to equations such as (2.8) in \ks.}
\eqn\defphi{u=e^{Q\phi\over2}.}
$Q$ is related to $k$ via
\eqn\Qk{Q^2={2\over k-2}\equiv-{2\over t}.}
The last equality defines
\eqn\defttt{t=-(k-2).}
$x$ is an auxiliary complex variable whose role
can be understood by expanding the operators $\Phi_h$ near the boundary
of $AdS_3$, $\phi\to\infty$, as is done on the second line
of \limphin. Note the difference
between the behavior for $h>1/2$ and $h<1/2$ \ks. For $h>1/2$,
the operators $\Phi_h$ are localized near the boundary at
$(\gamma,\bar\gamma)=(x,\bar x)$. For $h<1/2$, the delta function is
subleading, and the operators are smeared over the boundary. One can think of
$\Phi_h$ as the propagator of a particle with mass $h(h-1)$ from a point
$(x,\bar x)$ on the boundary to a point
$(\phi,\gamma,\bar\gamma)$ in the bulk of $AdS_3$. Thus, $x$ labels the
position on the boundary of $AdS_3$, which is the base space of the CFT
dual to string theory on $AdS_3$ via the AdS/CFT correspondence.

\lref\ZamolodchikovBD{
A.~B.~Zamolodchikov and V.~A.~Fateev,
``Operator Algebra And Correlation Functions In The Two-Dimensional
Wess-Zumino SU(2) X SU(2) Chiral Model,''
Sov.\ J.\ Nucl.\ Phys.\  {\bf 43}, 657 (1986)
[Yad.\ Fiz.\  {\bf 43}, 1031 (1986)].}

The operators $\Phi_h$ are primary under the $\widehat{SL(2)}$
current algebra \opealg; they satisfy
\eqn\rrr{\eqalign{
J^3(z) \Phi_h(x,\bar x;w,\bar w)\sim &-{(x\partial_x+h)\Phi_h(x,\bar x)
\over z-w}\cr
J^+(z) \Phi_h(x,\bar x;w,\bar w)\sim &-{\left(x^2\partial_x+2hx\right)
\Phi_h(x,\bar x)\over z-w}\cr
J^-(z) \Phi_h(x,\bar x;w,\bar w)
\sim &-{\partial_x\Phi_h(x,\bar x)\over z-w}~.\cr
}}
Their worldsheet scaling dimensions are
\eqn\scdim{\Delta_h=-{h(h-1)\over k-2}={h(h-1)\over t}~.}
It is very convenient \ZamolodchikovBD\ to ``Fourier transform'' the
$SL(2)$ currents as well, and define
\eqn\intoppp{J( x; z)\equiv -J^+(x; z)=2 x J^3(z)- J^+(z) - x^2 J^-(z).}
Since $J_0^-=-\partial_x$ is the generator of translations in $x$
(see \rrr) we can think of \intoppp\ as a result of ``evolving'' the
currents $J^a(z)$ in $x$:
\eqn\jofxdef{\eqalign{ J^+(x;z)=&e^{-xJ_0^-} J^+(z) e^{xJ_0^-}=J^+(z)
-2xJ^3(z)+x^2J^-(z)\cr
J^3(x;z)=&e^{-xJ_0^-} J^3(z) e^{xJ_0^-}=J^3(z) -xJ^-(z) =
-\half \partial_x J^+(x;z)\cr
J^-(x;z)=&e^{-xJ_0^-} J^-(z) e^{xJ_0^-}=J^-(z)
=\half \partial_x^2 J^+(x;z).\cr}}
The OPE algebras \opealg\ and \rrr\ can be written in terms of $J(x;z)$
as follows:
\eqn\JJ{J( x; z) J( y; w)\sim k {(y-x)^2\over( z- w)^2}+ { 1\over
z- w}\left[(y-x)^2\partial_y -2(y-x) \right] J( y; w)}
\eqn\JPhi{J( x; z) \Phi_h(y,\bar y;w,\bar w)\sim  {1 \over z-w} \left[(
y-x)^2\partial_y +2h(y-x)\right]\Phi_h(y, \bar y) .}

\noindent
It is sometimes useful to expand the operators \limphin\ in modes,
\eqn\ppp{\Phi_h(x,\bar x)=\sum_{m,\bar m}V_{h-1;m,\bar m} x^{-m-h}
\bar x^{-\bar m-h}}
or
\eqn\vjmbarm{V_{j;m,\bar m}=\int d^2x x^{j+m}\bar x^{j+\bar m}
\Phi_{j+1}(x,\bar x)~.}
Note that \rrr\ implies that $V_{j;m,\bar m}$ transforms under
$\widehat{SL(2)}$ as follows:
\eqn\vjmpm{\eqalign{
J^3(z)V_{j;m,\bar m}(w)=&{m\over z-w}
V_{j;m,\bar m}\cr
J^{\pm}(z)V_{j;m,\bar m}(w)=&{(m\mp j)\over z-w}
V_{j;m\pm 1,\bar m}.\cr}}

\noindent
The discussion above concerns the ``short string'' sector of the
model. The theory has other sectors, which contain long strings
located near the boundary of $AdS_3$. These are obtained by performing
spectral flow on the short string sector \mo\ (see also \ags). We will
not discuss the physics associated with long strings here, except for
some comments in section 3. It would be interesting to generalize the
results presented below to sectors with long strings. Since our analysis
is algebraic, it should be possible to obtain such results by performing
spectral flow \mo, or twisting as in \ags. Some results on correlation
functions including long strings appeared recently in \gn.

\lref\DiFrancescoUD{
P.~Di Francesco and D.~Kutasov,
``World sheet and space-time physics in two-dimensional (Super)string theory,''
Nucl.\ Phys.\ B {\bf 375}, 119 (1992)
[hep-th/9109005].
}
\lref\TeschnerRV{
J.~Teschner,
``Liouville theory revisited,''
hep-th/0104158.
}

\newsec{The Wakimoto representation and the spacetime central extension}

Consider the Lagrangian:
\eqn\Lwak{\CL=\partial\phi\bar\partial\phi-Q\widehat R\phi+
\beta\bar\partial\gamma+\bar\beta\partial\bar\gamma
-\lambda\beta\bar\beta e^{-Q\phi}~.}
Integrating out the fields $\beta, \bar\beta$ one obtains the
$AdS_3$ Lagrangian \lpoi, \defphi. The description \Lwak\ is useful
for studying the physics at large $\phi$ (near the boundary of
$AdS_3$). The interaction
term proportional to $\lambda$ goes to zero there, and one gets
a free linear dilaton theory for $\phi$, as well as a free
$(\beta,\gamma)$ system. Moreover, the effective string coupling
$g_s(\phi)\simeq\exp(-Q\phi/2)$ also goes to zero there, so the
system is weakly coupled in spacetime as well. All this should
be contrasted with the original Lagrangian \lpoi, which is
singular at $\phi\to\infty$.

Processes that are dominated by the large $\phi$ region can be studied
by viewing $(\phi,\beta,\gamma)$ as free fields with the propagators
\eqn\cnwsp{\langle\phi(z)\phi(0)\rangle=-\log|z|^2~, \qquad
\langle\beta(z)\gamma(0)\rangle={1\over z}}
and treating $\lambda$ perturbatively (as a screening charge).
This is a familiar technique in Liouville theory (see \eg\
\refs{\DiFrancescoUD,\TeschnerRV} for reviews); it has been applied
to $SL(2,R)$ CFT \eg\ in \refs{\gmoms,\BershadskyIN,\becker,\hos,\gn,\kazhos}.
Like in Liouville theory, generic correlation functions
cannot be studied this way, since they are sensitive to the region
$\phi\to 0$ (this is reasonable since the behavior of $g_s(\phi)$
mentioned above implies that interactions turn off as $\phi\to\infty$,
so particles have to penetrate to finite $\phi$ in order to interact).
Formally, one cannot expand in $\lambda$ in \Lwak\ since by shifting
$\phi$, $\lambda$ can be set to one.

Since we will be interested later in some situations were the physics
is dominated by the large $\phi$ region, we give next the form of some
of the objects described in the previous section in the free field
``Wakimoto variables'' $(\phi,\beta,\gamma)$. The current algebra is
represented by (normal ordering is implied):
\eqn\curalg{\eqalign{
J^3=&\beta\gamma+{1\over Q}\partial\phi~,\cr
J^+=&\beta\gamma^2+{2\over Q}\gamma\partial\phi
+k\partial\gamma~,\cr
J^-=&\beta~,\cr}}
where $Q$ is given in \Qk.

The $SL(2)$ primaries $\Phi_h$ behave (for $h>1/2$) like (see \limphin)
\eqn\phihh{\Phi_h\simeq -e^{Q(h-1)\phi}\delta^2(\gamma-x),}
or, performing the transform \vjmbarm,
\eqn\primr{V_{jm\bar m }=-\gamma^{j+m}\bar \gamma^{j+ \bar m}
e^{Qj\phi}~.}
The powers of $\gamma$ and $\bar \gamma$ can be both positive and
negative. The only constraint that follows from single valuedness on
$AdS_3$ is that $m-\bar m$ must be an integer. One can check directly
using free field theory that the scaling dimension of $V_{jm\bar m}$ is
\eqn\scvjm{\Delta(V_{jm\bar m})=-{j(j+1)\over (k-2)}~,}
in agreement with \scdim\ with
\eqn\hjpo{h=j+1~.}

\noindent
The coupling $\lambda$ in string theory on $AdS_3$ plays a role similar to
that of the cosmological constant $\mu$ in Liouville theory. For example,
the partition sum of the theory has a genus expansion of the form
\eqn\partsumexp{Z(\lambda, g_s)=\sum_{n=0}^\infty Z_n
\left({g_s^2\over \lambda}\right)^{n-1}~,}
where $Z_n$ is the genus $n$ partition sum\foot{For $n=1$ (the torus)
there is a logarithmic scaling violation.}.
Similarly, correlation functions of the operators $\Phi_h$ \limphin\
scale as (for simplicity and future use we exhibit the form for the
correlation functions on the sphere; similar formulae hold for
higher genus correlation functions):
\eqn\scspcor{\langle\Phi_{h_1}(x_1;z_1)\cdots\Phi_{h_n}(x_n;z_n)\rangle
=\lambda^sF_n(x_1,\cdots, x_n;z_1,\cdots,z_n)~,}
where
\eqn\ssvalue{s=1+\sum_{i=1}^n(h_i-1)~,}
and $F_n$ contains the non-trivial information about the correlation
function. We see that, as mentioned above, the physics is essentially
independent of $\lambda$ -- it can be absorbed into $\phi$ (\ie\ it can be
set to one by rescaling $g_s$ and the operators). Nevertheless,
as in Liouville theory, for some purposes it is convenient to
keep the $\lambda$ dependence explicit.

It is natural to ask what is the invariant meaning of the Wakimoto
coupling $\lambda$ in string theory on $AdS_3$? In other words,
can one describe the Wakimoto interaction in \Lwak\ as an observable
in the theory in a parametrization independent way? As we have just
seen, changing $\lambda$ changes the effective string coupling of
the model \partsumexp. Therefore, it is natural to expect that the
operator that changes $\lambda$ should be the zero momentum mode
of the dilaton. It is in fact well known that the dilaton is massive
in string theory on $AdS_3$, but its zero mode is tunable. The
corresponding vertex operator was constructed and discussed
in \ks; its form is (see eq. (4.18) in \ks)
\eqn\iii{I=-{1\over k^2}\int d^2z J(x;z)\bar J(\bar x;\bar z)
\Phi_1(x,\bar x;z,\bar z)~.}
This operator has many special properties \ks. It commutes with
the full $SL(2)_L\times SL(2)_R$ affine Lie algebra; this
follows from the OPE
\eqn\opeJJPhi{J(x;z)\left[J(y;w)\Phi_1(y,\bar y; w,\bar w)\right]\sim
k(x-y)^2\partial_w\left[{\Phi_1(y, \bar y; w, \bar w)\over
z-w}\right]~,}
and its right moving analog.
The operator $J(x;z)\bar J(\bar x;\bar z)\Phi_1(x,\bar x;z,\bar z)$ is
marginal on the worldsheet. From the point of view of the spacetime CFT,
$I$ is a dimension zero operator (but, as
we will see soon it is {\it not} a multiple of the identity operator).
It can be shown to satisfy $\partial_x I=\partial_{\bar x} I=0$; hence
it is constant in correlation functions (but not necessarily the same
constant in different correlation functions).

Therefore, it is natural to conjecture that changing $\lambda$ in \Lwak\
corresponds to adding to the worldsheet action the operator $I$ \iii.
To see that this makes sense qualitatively, we next show  that the large
$\phi$ dependence of $I$ agrees with the Wakimoto interaction term in \Lwak.

The leading behavior of $\Phi_1(x,\bar x;z,\bar z)$ as $\phi\to\infty$
is given by \limphin:
\eqn\iik{\Phi_1\simeq-\delta^2(\gamma-x)
-{e^{-Q\phi}\over\pi|\gamma-x|^4}~.}
As for $J(x;z)$, its behavior can be read off eq. \curalg\ together
with the definition \intoppp:
\eqn\iil{J(x;z)=-\beta(x-\gamma)^2+{2\over Q}(x-\gamma)\partial\phi-
k\partial\gamma~.}
We would now like to analyze the large $\phi$ behavior of
$J(x;z)\bar J(\bar x;\bar z)\Phi_1$. The first term in \iik\
(the $\delta^2(x-\gamma)$) contributes only
\eqn\igggg{I=\int d^2z
\partial\gamma\bar\partial\bar\gamma\delta^2(x-\gamma)+...~.}
This is a vacuum contribution; the ``...'' stand for subleading
contributions. As explained in \refs{\gks,\ks}, \igggg\
measures the number of long strings in the vacuum.

The screening charge in \Lwak\ should thus
come from the first subleading contribution,
where we take the second term from
$\Phi_1$ \iik\ and multiply by \iil. Since, as proven in \ks, the
operator $I$ is independent of $x$, $\bar x$, the only term in the
product $J(x;z)\bar J(\bar x;\bar z)\Phi_1$ that contributes to
correlation functions is the first one in \iil\ (which can be seen
\eg\ by plugging \iik, \iil\ in \iii\ and taking $x\to\infty$);
thus, the first subleading contribution to $I$ is
\eqn\iim{I\simeq\int d^2z
\partial\gamma\bar\partial\bar\gamma\delta^2(x-\gamma)
+{1\over\pi k^2}\int d^2z\beta\bar\beta
e^{-Q\phi}~,}
in agreement with the Wakimoto screening charge \Lwak.

To establish the precise relation between $\lambda$ in \Lwak\ and
a worldsheet deformation by $I$ of the form
\eqn\iideform{\SS(\rho)=\SS_0+\rho I~,}
one can proceed as follows.
Consider the partition sum
\eqn\zzrho{Z(\rho)=\langle e^{-S(\rho)}\rangle~.}
As we will see shortly, on the sphere one has
\eqn\iicorsph{\langle e^{-\rho I}\rangle=Z_0e^{-\rho\langle I\rangle}~.}
Here and below $\langle I\rangle$ is the one point function
of $I$ on the sphere normalized by dividing it by the partition sum
on the sphere, unless stated otherwise.
Comparing \iicorsph\ to the tree level term in \partsumexp\
we see that the relation between $\lambda$ in \Lwak\ and $\rho$ in
\iideform\ is
\eqn\rholamrel{\lambda=e^{-\rho\langle I\rangle}~,}
where we normalized $\lambda$ such that $\rho=0$ corresponds to $\lambda=1$.
Note that \rholamrel\ is a natural relation; $\rho$ changes
the expectation value of the zero mode of the dilaton (see \ks),
whereas $\lambda$ is proportional to $g_s^{-2}$ (see \partsumexp).
Thus, it is natural that $\rho$ goes like $\log\lambda$.

It remains to establish \iicorsph, but before getting to that
note that combining \rholamrel\ with \scspcor,
\iideform\ and \zzrho\ leads to another interesting relation.
Differentiating \scspcor\ once w.r.t. $\rho$ gives:
\eqn\irelcor{\langle I\Phi_{h_1}\cdots\Phi_{h_n}\rangle
=\left(1+\sum_{i=1}^n(h_i-1)\right)\langle I\rangle\langle
\Phi_{h_1}\cdots\Phi_{h_n}\rangle~.}
We will next prove \iicorsph\ and \irelcor\ by a direct
calculation, thereby establishing the correspondence \rholamrel.

Consider first the relation \irelcor. To calculate the l.h.s. one
notes \ks\ that the operator $I$ \iii\ is ``almost'' a total derivative.
In fact, one has
\eqn\jphitotder{\bar J\Phi_1=-{k\over\pi}\bar\partial\Lambda~,}
where $\Lambda$ is given explicitly in \ks; we will only need
its large $\phi$ form,
\eqn\largephilam{\lim_{\phi\to\infty}\Lambda={1\over x-\gamma}~.}
Plugging \jphitotder\ into \iii\ and using the fact that
$J(x;z)$ is holomorphic (up to contact terms), we find that
\eqn\iitotder{I={1\over \pi k}\int d^2z\bar\partial(J\Lambda)~.}
Despite appearances, \iitotder\ does not imply that $I$ is trivial.
The technical reason for that is that $\Lambda$ is not a good
observable on $AdS_3$ (see \ks\ for a more detailed discussion).
In particular, it transforms like a primary with $h=0$ under
$SL(2)_R$ (see \JPhi) but its transformation as an object
with $h=1$ under $SL(2)_L$ contains an anomalous term
\eqn\JLambda{J( x; z) \Lambda(y,\bar y;w,\bar w)\sim  {\left[( y-
x)^2\partial_y+2(y-x)\right]\Lambda(y, \bar y;w,\bar w)-1\over z-w}~.}
Therefore, the composite operator $J(x;z)\Lambda(x;z)$ requires
normal ordering, and should be defined via a limiting procedure
\eqn\norord{J(x;z)\Lambda(x;z)=
\lim_{z\to z'} J(x;z)\Lambda(x;z')+{1\over(z-z')}~.}
In addition, the operator $J\Lambda$ in \iitotder\ has short distance
singularities when it approaches other operators, which also give
contributions to correlators involving $I$. From the spacetime point of
view, this happens because $\Lambda$ is associated with a gauge
transformation that does not go to zero at infinity (the boundary
of $AdS_3$).

At any rate, returning to \iitotder, we conclude that the correlator
$\langle I\Phi_{h_1}\cdots\Phi_{h_n}\rangle$ receives contributions
from the boundaries of moduli space, which are small circles around
the insertions $z_i$ (as described in detail in \ks), and an
additional contribution $(1/k)\langle\Phi_{h_1}\cdots\Phi_{h_n}\rangle$
from the anomalous term in \norord.
To compute the contributions from the small circles around the insertions
one uses the OPE algebra \JPhi\ and the relation (see \refs{\ks,\tesss})
\eqn\opeLPhi{\lim_{z_1\rightarrow z_2}
\Lambda(x_1,\bar x_1; z_1, \bar z_1) \Phi_h(x_2,\bar x_2; z_2, \bar z_2)
= {1\over x_1-x_2} \Phi_h(x_2,\bar x_2; z_2, \bar z_2)~.}
This leads to:
\eqn\Ihhhh{\langle I\Phi_{h_1}\cdots\Phi_{h_n}\rangle=
{1\over k}\left(1+\sum_{i=1}^n(h_i-1)\right)
\langle\Phi_{h_1}\cdots\Phi_{h_n}\rangle~.}
Comparing to \irelcor\ we see that the structure is the same if
\eqn\ivev{\langle I\rangle={1\over k}~.}
The result \ivev\ can be proven directly as a part of the derivation of
\iicorsph, to which we turn next. Consider the correlator
\eqn\corIII{\II_n=\langle I^n\rangle~.}
$\II_n$ can be computed using the same logic as before.
Rewrite one of the $n$ insertions of $I$ as \iitotder; there are again
potential boundary contributions from small circles around the other
insertions, and a contribution from the anomalous term in
\norord. In this particular case, the contributions from the
vicinity of the other insertions vanish. This is not difficult
to see using the OPE \opeJJPhi.

Hence, the analog of \Ihhhh\ for this case is
\eqn\irecrel{\II_n={1\over k}\II_{n-1}~.}
This is equivalent to \iicorsph\ with the expectation value
of $I$ given by \ivev.

To summarize, we have proven the relation \rholamrel, by establishing
its consequences \iicorsph\ and \irelcor.

The preceeding discussion has interesting consequences for the
structure of the spacetime Virasoro (and affine) algebras
in string theory on $AdS_3$. As shown in \ks, the spacetime
stress tensor in string theory in $AdS_3$, $T(x)$, is given by a
certain integrated vertex operator. $T(x)$ satisfies the spacetime
Virasoro OPE
\eqn\virope{T(x)T(y)\sim{c_{\rm st}/2\over(x-y)^4}+{2T(y)\over (x-y)^2}
+{\partial T(y)\over x-y},}
where the central term is given by
\eqn\cst{c_{\rm st}=6kI~.}
Since in all known two dimensional CFT's the central charge is proportional
to the identity operator, it was assumed in \ks\ that this is the case
here as well. We now see from \irelcor\ that the operator
$I$ appears to be a non-trivial dimension zero operator. Thus, string
theory on $AdS_3$ (with NS background) has the property that the central
extension is not proportional to the identity operator.

\lref\deBoerPP{
J.~de Boer, H.~Ooguri, H.~Robins and J.~Tannenhauser,
``String theory on AdS(3),''
JHEP {\bf 9812}, 026 (1998)
[hep-th/9812046].
}

This interesting behavior seems to be
directly related to the non-locality of the theory.
It is well known that in string theory on $AdS_3$ long strings
carry central charge (see \eg\ \refs{\gks,\ks}). Each long string
carries central charge $c=6k$. This is in fact one of the consequences
of \Ihhhh. More generally, \Ihhhh\ predicts that ``short strings'' carry
central charge as well. Indeed, consider a short string state
$|h\rangle$, created by acting with the local operator $\Phi_h$
on the vacuum. The central charge \cst\ in this state is given by
\eqn\csthh{c_{\rm st}(h)=6k
{\langle \Phi_h I \Phi_h\rangle\over\langle \Phi_h\Phi_h\rangle}~.}
One type of contribution comes from disconnected worldsheets \deBoerPP,
with the operator $I$  on one worldsheet and the two $\Phi_h$ operators
on a second worldsheet. These give vacuum contributions to the spacetime
central charge,
\eqn\cstvacu{c_{\rm st}^{({\rm vac})}=2(x-y)^4\langle T(x)T(y)\rangle=
6k\langle I\rangle.}
Here $\langle I\rangle$ is the one point function of $I$, computed without
dividing by the partition sum on the sphere (unlike eqs. like \ivev).
This vacuum central charge has an expansion in powers of $g_s$, but it is
clearly independent of $h$ (by definition).

The leading correction to this vacuum contribution comes from
the amplitude where all three operators in \csthh\ are on the
same worldsheet of spherical topology. Using \Ihhhh\ this gives
\eqn\cstexp{c_{\rm st}(h)=c_{\rm st}^{({\rm vac})}+6(2h-1)~.}
Thus, we see that the central charge increases with $h$.
When $h$ reaches $h\sim k/2$, the threshold for creating long
strings, one finds that the central charge carried by the state
is $c\simeq 6k$, in agreement with the long string picture.
The result of \refs{\gks,\ks}, that long strings carry $c_{\rm st}=6k$
can be shown as follows. A long string vertex operator is obtained
by taking a short string vertex operator from the continuous
series, with $h=\half+i\lambda$, and applying to it a certain
twist operator described in \ags\ (which implements the spectral
flow of \mo). The twist field contributes $+1$ to $I$ due to the
presence of the first term in \iim. The contribution of the vertex
operator from the continuous series is given by \Ihhhh:
\eqn\contcont{
\langle I\Phi_{\half+i\lambda}\Phi_{\half-i\lambda}\rangle=
{1\over k}\left(1+(-\half+i\lambda)+(-\half-i\lambda)\right)
\langle\Phi_{\half+i\lambda}\Phi_{\half-i\lambda}\rangle=0.}
Therefore, we conclude that long string states with winding
number one carry $c_{\rm st}=6k$, as expected.

\lref\TeschnerYF{
J.~Teschner,
``On the Liouville three point function,''
Phys.\ Lett.\ B {\bf 363}, 65 (1995)
[hep-th/9507109].
}

\newsec{Degenerate representations and correlation functions}

In this section we compute the OPE coefficients of certain degenerate
operators in CFT on $AdS_3$ with other primaries $\Phi_h$ \limphin.
This can be done by using the conjectured duality between the $SL(2)/U(1)$
and Sine-Liouville CFT's \fzzsl; the results can be compared to a direct analysis
in $SL(2)$ CFT, which we review in section 5. As we will see, the two
approaches agree, providing a test of the duality.

The above OPE coefficients can be used to
compute correlation functions in $SL(2)$ CFT,
following a method used in Liouville theory \TeschnerYF.
As an example, we calculate the two point function
\eqn\twoptfn{\langle \Phi_h(x,\bar x;z,\bar z)
\Phi_{h'}(y,\bar y;w,\bar w)\rangle=\delta(h-h')
{D(h)\over |x-y|^{4h}|z-w|^{4\Delta_h}}~,}
which was obtained before in \refs{\tesch,\fzzsl}.
The duality provides an efficient way for calculating
\twoptfn\ and other correlation functions.

The degenerate operators in $SL(2)$ CFT are of the form
\limphin\ with (see \tesch\ for more details):
\eqn\degops{h_{r,s}={1-r\over2}-{1-s\over2}t~,\qquad r,s=1,2,3,\dots~.}
For irrational $k$, the Fock module corresponding to $h_{r,s}$
contains a single null state at level $r(s-1)$. Consider, for example,
the special case $s=1$. The degenerate
representations have $h_{r,1}=(1-r)/2$,
and the null state is at level zero.
Looking back at \limphin\ this is natural: $\Phi_{h_{r,1}}$
is in this case a polynomial of degree $r-1$ in $x$ and $\bar x$,
and the null state is
\eqn\nullrsz{\partial_x^r\Phi_{(1-r)/2}=
\partial_{\bar x}^r\Phi_{(1-r)/2}=0~, \qquad r=1,2,3,\dots~.}
The operators $\Phi_{(1-r)/2}$ correspond to finite, $r$ dimensional,
representations of $SL(2)$. They are direct generalizations of the
finite dimensional spin $(r-1)/2$ representations of $SU(2)$ which
are described in the language used here in \ZamolodchikovBD.

It should be emphasized that the fact that the quantum operators
$\Phi_{(1-r)/2}$ satisfy the null state equations \nullrsz\ is not
completely obvious. It is certainly true that they satisfy these
equations semiclassically, and that if the combination on the l.h.s.
of \nullrsz\ is not zero, it is a new $\widehat{SL(2)}$ primary in
the theory, whose norm is zero. In a unitary CFT this would mean that
this new primary vanishes, but $SL(2)$ CFT is not unitary, and so this
argument does not apply. It seems that the fact that the null states associated
with the degenerate representations \degops\ vanish is part of the
definition of $SL(2)$ CFT.

As we will see, to calculate $D(h)$ in \twoptfn\
it is enough to consider two degenerate operators:
\item{(a)}
The first non-trivial~\foot{The operator $\Phi_{h_{1,1}}$ is the identity.}
operator in the series $\Phi_{h_{r,1}}$:
\eqn\phitwone{\Phi_{h_{2,1}}\equiv \Phi_{-\half}~.}
\item{(b)}
The first non-trivial operator in the series $\Phi_{h_{1,s}}$:
\eqn\phionetwo{\Phi_{h_{1,2}}\equiv \Phi_{t\over 2}~.}

\noindent
We next discuss these two cases.

\subsec{$\Phi_{-\half}$}

Consider first the case $h=-1/2$. As mentioned above,
a look at \limphin\ makes it
clear that in this case the operator $\Phi_h$ reduces to a finite
polynomial,
\eqn\degone{
\Phi_{-\half}={2\over\pi}\left(|\gamma-x|^2e^{Q\phi\over2}+
e^{-{Q\phi\over2}}\right)~.}
Thus, the mode expansion \ppp\ is very simple:
\eqn\modehalf{\Phi_{-\half}=\sum_{m,\bar m=-\half,\half}
V_{-\frac32;m,\bar m}x^{\half-m}\bar x^{\half-\bar m}~,}
with
\eqn\modeshalf{\eqalign{
V_{-\frac32;\half,\half}=&{2\over\pi}\gamma\bar\gamma e^{Q\phi\over2}~,\cr
V_{-\frac32;-\half,-\half}=&{2\over\pi} e^{Q\phi\over2}~,\cr}}
etc. Note that, following standard practice, we have written in
\modeshalf\ only the large $\phi$ forms of the vertex operators.

There is actually an interesting subtlety here that will be important
later. The formula \degone\ from which \modeshalf\ follows is a
semiclassical expression, which should be valid in the $t\to-\infty$
limit. It is possible that for finite $t$ it will receive finite
renormalization, and this turns out indeed to be the case. To
determine this finite renormalization we proceed as follows.
We use the following two pieces of information:

\item{(1)} As $\phi\to\infty$, the operators $\Phi_h$ with $h>1/2$
behave as \limphin:
\eqn\phbeas{\Phi_h\simeq
-e^{Q(h-1)\phi}\delta^2(\gamma-x)~.}
The fact that the coefficient of the exponential is $-1$ is a choice
of the normalization of $\Phi_h$ with $h>1/2$ in the quantum theory.
\item{(2)} Once the normalization of the operators with $h>1/2$ is chosen,
there is no further freedom. The normalization of the operators $\Phi_h$
with $h<1/2$ is fixed, since they are related to those
with $h>1/2$ by a reflection symmetry described in \tesch:
\eqn\teschone{\Phi_h(x;z)=\RR(h){2h-1\over\pi}\int d^2x'
|x-x'|^{-4h}\Phi_{1-h}(x';z)~,}
where the $x'$ integral runs over the plane. We will see later
that the reflection coefficient $\RR(h)$ depends on the coupling
$\lambda$ \Lwak. For the value of $\lambda$ that we will be using
below, it is equal to
\eqn\rrhh{\RR(h)={\Gamma(1+{2h-1\over t})\over \Gamma(1-{2h-1\over t})}~.}

\noindent
Using points (1), (2) above, we can determine the $\phi\to\infty$
behavior of $\Phi_{-\half}$ by using the fact that
\eqn\behthhf{\Phi_{3\over2}(x)\simeq -e^{\half Q\phi}\delta^2(\gamma-x)~.}
Plugging this into \teschone, we find that
\eqn\behhalf{\Phi_{-\half}\simeq{2\over\pi}\RR(-\half)|x-\gamma|^2
e^{\half Q\phi}~.}
Comparing to the semiclassical expression \degone, we see that the
quantum correction is a multiplicative factor of
\eqn\ronehalf{\RR(-\half)={\Gamma(1-{2\over t})\over\Gamma(1+{2\over t})}~.}
We see that in the classical limit $t\to-\infty$ it goes to $1$, as
expected.

$\Phi_{-\half}$ satisfies the differential equations \nullrsz:
\eqn\degeq{\partial_x^2\Phi_{-\half}=\partial_{\bar x}^2\Phi_{-\half}=0~.}
Equation \degeq\ places strong constraints on the OPE
algebra of $\Phi_{-\half}$ with other operators. Indeed, consider
the OPE
\eqn\phitwoone{\Phi_{-\half}(x)\Phi_h(y)=
\sum_{h'} C_{hh'}|x-y|^{2(\half-h+h')}\Phi_{h'}(y)+\cdots~.}
The form of the r.h.s. is determined by spacetime conformal
invariance of string theory on $AdS_3$, and the ``$\cdots$'' stands
for contributions of descendants (under spacetime and worldsheet
Virasoro). Also, we have suppressed the dependence on the worldsheet
locations of  the operators. $C_{hh'}$ are structure constants
that need to be determined.

The differential equations \degeq\ place the following constraint
on the r.h.s. of \phitwoone:
\eqn\degrepc{C_{hh'}(\half-h+h')(-\half-h+h')=0~.}
Thus, there are only two possible terms in the sum on the r.h.s.
of \phitwoone, corresponding to $h'=h-\half$ and $h'=h+\half$:
\eqn\opehalf{\Phi_{-\half}(x)\Phi_h(y)=
C_-(h)\Phi_{h-\half}(y)+|x-y|^2C_+(h)\Phi_{h+\half}(y)+\cdots~.}
We would like next to determine the structure constants
$C_\pm$.

Experience with Liouville theory and Feigin-Fuchs representations
of various rational CFT's leads one to expect that
the physics of the degenerate operator $\Phi_{-\half}$ should
be {\it perturbative} in some description of the theory. In the
present context this means that the physics of degenerate
operators is dominated by the region $\phi\to\infty$, \ie\
the vicinity of the boundary of $AdS_3$. As far as we know,
the origin of this phenomenon is not well understood but,
as we will see, one can use it to determine the structure
constants in \opehalf.

We will next show that OPE's which involve $\Phi_{-\half}$
are perturbative in the Wakimoto variables described in section 3.
Consider first the free theory obtained by setting the Wakimoto
coupling $\lambda$ in \Lwak\ to zero. This already gives rise to
the second term in \opehalf. Indeed, at large $\phi$ we have
(see \limphin, \behhalf; we are assuming that $h>\half$)
\eqn\expphi{\eqalign{
&\Phi_{-\half}(x)\simeq {2\over\pi}\RR(-\half)|\gamma-x|^2e^{Q\phi\over2}~,\cr
&\Phi_h(y)\simeq -e^{Q(h-1)\phi}\delta^2(\gamma-y)~.\cr
}}
Multiplying the two, using the fact that $\gamma(z)$ behaves as a c-number
in these calculations, gives
\eqn\aaaa{\eqalign{
\Phi_{-\half}(x)\Phi_h(y)\simeq &-{2\over\pi}\RR(-\half)
|x-y|^2e^{Q(h-\half)\phi}\delta^2(\gamma-y)+\cdots=\cr
&{2\over\pi}\RR(-\half)|x-y|^2\Phi_{h+\half}(y)+\cdots~.\cr}}
Comparing to \opehalf\ we see that
\eqn\cplus{C_+={2\over\pi}\RR(-\half)~.}
The fact that the OPE \aaaa\ does not involve $\lambda$ implies that
the relevant interaction occurs very far from the ``wall'' provided
by $\lambda$, \ie\ it is a ``bulk interaction'' in the sense of
\DiFrancescoUD. It can occur anywhere in the infinite region near
the boundary of $AdS_3$.

It remains to compute $C_-$. To do that, \eg\ set $x=0$, which gives
(see \modehalf)
\eqn\cminus{\RR(-\half)V_{-\frac32;\half,\half}\Phi_h(y)=
C_-\Phi_{h-\half}(y)+\cdots~.}
For simplicity, set also $y=0$ (this gets rid of contributions from
descendants). We now have
\eqn\cca{{2\over\pi}\RR(-\half)\gamma\bar\gamma e^{\half Q\phi(z_1)}
e^{Q(h-1)\phi(z_2)}\delta^2(\gamma(z_2))=C_-
e^{Q(h-\frac32)\phi(z_2)}\delta^2(\gamma(z_2))+\cdots~.}
We see that in this case the $\phi$ charges do not add up
correctly, but if we bring down from the action \Lwak\ one power of the
interaction, we seem to land on our feet:
\eqn\ccb{\eqalign{
&{2\lambda\over\pi}\RR(-\half)
\int d^2z\beta\bar\beta(z)e^{-Q\phi(z)}
\gamma\bar\gamma(z_1) e^{Q\phi(z_1)\over2}
e^{Q(h-1)\phi(z_2)}\delta^2(\gamma(z_2))=\cr
&C_-
e^{Q(h-\frac32)\phi(z_2)}\delta^2(\gamma(z_2))+\cdots~.\cr}}
The exponentials of $\phi$ simply give
\eqn\ccc{|z-z_1|^{Q^2}|z-z_2|^{2Q^2(h-1)} e^{Q(h-\frac32)\phi(z_2)}~.}
Using the fact that
\eqn\ccd{\beta(z) f(\gamma(z_2))\simeq{1\over z-z_2} f'(\gamma(z_2))~,}
for any function $f$, we have
\eqn\cce{\eqalign{
&\beta\bar\beta(z)\gamma\bar\gamma(z_1)\delta^2(\gamma(z_2))=\cr
&\left({1\over z-z_1}+{1\over z-z_2}\gamma(z_1){\partial\over\partial
\gamma(z_2)}\right)
\left({1\over \bar z-\bar z_1}+{1\over
\bar z-\bar z_2}\bar\gamma(\bar z_1){\partial\over\partial
\bar\gamma(\bar z_2)}\right)\delta^2(\gamma(z_2))~.\cr}}
In the $\gamma{\partial\over\partial\gamma}$ terms we can set
$z_1=z_2$ since we are only interested in the most singular
terms as $z_1\to z_2$. This allows one to integrate by parts so
\eqn\ccf{\eqalign{
&\beta\bar\beta(z)\gamma\bar\gamma(z_1)\delta^2(\gamma(z_2))=\cr
&\left({1\over z-z_1}-{1\over z-z_2}\right)
\left({1\over \bar z-\bar z_1}-{1\over \bar z-\bar z_2}\right)
\delta^2(\gamma(z_2))={|z_1-z_2|^2\over |z-z_1|^2|z-z_2|^2}
\delta^2(\gamma(z_2))~.\cr}}
We see that the r.h.s. of \ccc\ and \ccf\ is exactly of the form
expected in \cca; hence
\eqn\ccg{C_-={2\lambda\over\pi}\RR(-\half)
\int d^2z|z-1|^{Q^2-2}|z|^{2Q^2(h-1)-2}~.}
Using standard results we get
\eqn\cch{\eqalign{
C_-(h)=&2\lambda\RR(-\half){\Gamma(\half Q^2)\Gamma(Q^2(h-1))\Gamma(1-\half
Q^2(2h-1))\over \Gamma(1-\half Q^2)\Gamma(1-Q^2(h-1))\Gamma(\half
Q^2(2h-1))}\cr
=&
2\lambda\RR(-\half){\Gamma(-{1\over t})\Gamma(-{2(h-1)\over t})\Gamma(1+
{2h-1\over t})\over \Gamma(1+{1\over t})\Gamma(1+{2(h-1)\over t})
\Gamma(-{2h-1\over t})}~.\cr}}
Physically, the analysis above means that the interaction implied
by the first term on the r.h.s. of \opehalf\ is also a ``bulk interaction''
but it involves in addition to the fields $\Phi_{-\half}$, $\Phi_h$ and
$\Phi_{h-\half}$ that appear explicitly also a zero momentum dilaton,
whose vertex operator is given by the Wakimoto screening charge, as discussed
in section 3.

This concludes the calculation of $C_\pm$ in \opehalf. We can now use these
structure constants to obtain a constraint on the correlators of $\Phi_h$.
Consider \eg\ the three point function
\eqn\thpt{\langle\Phi_{-\half}(x)\Phi_h(y_1)\Phi_{h+\half}(y_2)\rangle~.}
If we first send $x\to y_1$ and use the $C_+$ term in \opehalf\ we get
\eqn\twoone{C_+(h)D(h+\half)~.}
Taking $x\to y_2$ first gives
\eqn\twotwo{C_-(h+\half)D(h)~.}
Equating \twoone, \twotwo\ we conclude that
\eqn\zzxx{C_-(h+\half)D(h)=C_+(h) D(h+\half)~.}
Shifting $h$ by $1/2$ for convenience we conclude that
\eqn\ttt{{D(h)\over D(h-\half)}={C_-(h)\over C_+(h-\half)}=
\pi\lambda{\Gamma(-{1\over t})\over\Gamma(1+{1\over t})}
{\Gamma(-{2(h-1)\over t})\over\Gamma(1+{2(h-1)\over t})}
{\Gamma(1+{2h-1\over t})\over\Gamma(-{2h-1\over t})}~.}
One class fo solutions (which turns out to be the relevant
one, as we show below) is
\eqn\corans{D(h)=\nu^{2h-1}
{\Gamma(1+{2h-1\over t})\over\Gamma(-{2h-1\over t})}~,}
where $\nu$ is a constant (function of $k$ and $\lambda$):
\eqn\nulagg{\nu=
\pi\lambda{\Gamma(-{1\over t})\over\Gamma(1+{1\over t})}~.}
The result of Teschner \tesch\ is the same as \corans\ with
\eqn\nnnuuu{\nu={\Gamma(1-{1\over t})\over\Gamma(1+{1\over t})}~.}
By tuning $\lambda$ (to $1/\lambda=-t\pi$), the two results agree.

In any case, as is clear from the discussion above, $\lambda$
plays here a similar role to $\mu$ in Liouville theory, \ie\ changing
$\lambda$ rescales the operators $\Phi_h$ by an $h$-dependent factor.
A nice choice of normalization that we will use below is
\eqn\lamval{\lambda=
{1\over\pi}{\Gamma(1+{1\over t})\over\Gamma(-{1\over t})}~,}
which eliminates the factor $\nu$ from the two point function
\corans\ (\ie\ sets $\nu=1$ in \nulagg).

\subsec{$\Phi_{t\over2}$}

We now move on to a discussion of the second degenerate operator,
\phionetwo. One way to introduce it is the following. Consider
operators of the form (normal ordering implied)
\eqn\abc{\theta(x)=a\partial_x^2 J \Phi_h+b\partial_xJ\partial_x\Phi_h
+c J\partial_x^2\Phi_h~.}
It is clearly a descendant of $\Phi_h$, but for some combination of
the numbers $a,b,c,h$ it might be a primary again.
To find $a,b,c,h$ it is convenient to use the language of states.
Using \jofxdef\ and the fact that
\eqn\dda{\Phi_h(x)=e^{-xJ_0^-}\Phi_h(0)e^{xJ_0^-}}
(see \eg\ \ks\ eq. (2.9)), and
\eqn\ddb{|h\rangle=\Phi_h(0)|0\rangle~,}
we have
\eqn\ddc{|\theta\rangle=\left[-2aJ_{-1}^--2bJ_{-1}^3J_0^--
cJ_{-1}^+(J_0^-)^2\right]|h\rangle~.}
We would like to check that
\eqn\ddd{J_n^a|\theta\rangle=0~,\;\;\;\forall\;n\ge 1~,\;\;\; a=3,\pm~.}
Only $J_1^a$ give non-trivial constraints.
Applying $J_1^-$ and using
\eqn\dde{\eqalign{
&[J_1^-, J^3_{-1}]=J_0^-~,\cr
&[J_1^-, J^+_{-1}]=2J_0^3+k~,\cr}}
we have
\eqn\ddf{J_1^-|\theta\rangle=
\left[-2b-c(k+2(-h-2))\right](J_0^-)^2|h\rangle~.}
Hence,
\eqn\ddg{2b+c[k-2(h+2)]=0~.}
Applying $J_1^3$ and using
\eqn\ddg{\eqalign{
&[J_1^3, J^\pm_{-1}]=\pm J_0^\pm~,\cr
&[J_1^3, J^3_{-1}]=-\half k~,\cr}}
we have
\eqn\ddh{J_1^3|\theta\rangle=\left[2aJ_0^-+2b\half
kJ_0^--cJ_0^+(J_0^-)^2\right]|h\rangle~.}
Since $J_0^+|h\rangle=0$, we can commute it through, using
$$[J_0^+, J_0^-]=-2J_0^3~.$$
Collecting all the terms we get
\eqn\ddi{2a+kb-2c(2h+1)=0~.}
Similarly, applying $J_1^+$ and using
\eqn\ddii{\eqalign{
&[J_1^+, J^3_{-1}]=-J_0^+~,\cr
&[J_1^+, J^-_{-1}]=-2J_0^3+k~,\cr}}
we have
\eqn\ddiii{J_1^+|\theta\rangle=
\left[-2a(k-2J_0^3)-2b(-J_0^+)J_0^-\right]|h\rangle~.}
Rearranging terms, we conclude that
\eqn\ddj{4bh-2a(k+2h)=0~.}
The equations \ddg, \ddi, \ddj\ have a unique solution
(up to rescaling $a,b,c,$ which is of course a symmetry
of \abc, and up to $h\to 1-h$ which is the reflection symmetry,
\teschone)
\eqn\ddk{\eqalign{
a&=\half t(t+1)\cr
b&=t+1\cr
c&=1\cr
h&={t\over 2}~.\cr
}}
Since \abc, \ddk\ is a primary of $\widehat{SL(2)}$ which
is also a current algebra descendant, and its norm is zero,
it is natural to set it to zero.
Again, as in the discussion of $\Phi_{-\half}$
above, this is not obvious, since the theory is not unitary.
In this case, one cannot even verify semiclassically that
the combination \abc\ is zero, since the operator does not have
a smooth $t\to-\infty$ limit. Nevertheless, we will assume
that the null state should be set to zero as part of the definition
of the theory.

Requiring that $\theta=0$ imposes constraints on the OPE
\eqn\opettwo{\Phi_{t\over2}(x)\Phi_h(y)=
\sum_{h'}|x-y|^{-2(\half t+h-h')}C_{hh'}\Phi_{h'}(y)+\cdots~.}
We must have
\eqn\ooa{0=\theta(x)\Phi_h(y)=
\half t(t+1)\partial_x^2 J\Phi_{t\over2}(x)\Phi_h(y)
+(t+1)\partial_x J\partial_x\Phi_{t\over2}\Phi_h(y)+
J(x)\partial_x^2\Phi_{t\over2}(x)\Phi_h(y)~.}
Using the OPE algebra \JPhi, we have:
\eqn\oob{\eqalign{
&t(t+1)\Phi_{t\over2}(x)\partial_y\Phi_h(y)+\cr
&(t+1)\partial_x\Phi_{t\over2}(x)\left[2(x-y)\partial_y-2h\right]
\Phi_h(y)+\cr
&\partial_x^2\Phi_{t\over2}(x)\left[(y-x)^2\partial_y
+2h(y-x)\right]\Phi_h(y)=0~.\cr}}
Substituting \opettwo\ into \oob\ we conclude that
\eqn\ooc{\eqalign{
&t(t+1)\partial_y(x-y)^{-{t\over2}-h+h'}+\cr
&(t+1)\left[2(x-y)\partial_y-2h\right]\partial_x(x-y)^{-{t\over2}-h+h'}+\cr
&\left[(y-x)^2\partial_y
+2h(y-x)\right]\partial_x^2(x-y)^{-{t\over2}-h+h'}=0~.\cr
}}
This gives a cubic equation for $h'$, whose solutions are:
\eqn\ood{h'=h\pm {t\over2},\;\;1-{t\over2}-h~.}
To summarize, we find that
\eqn\ooe{\Phi_{t\over2}(x)\Phi_h(y)=C_1(h)\Phi_{h+{t\over2}}(y)+
C_2(h)|x-y|^{-2t}\Phi_{h-{t\over2}}(y)+C_3(h)|x-y|^{2(-t-2h+1)}
\Phi_{-{t\over2}-h+1}~.}
We will next show that the problem of determining
$C_1$, $C_2$, $C_3$ is perturbative in the
Sine-Liouville description. First we need to find the
large $\phi$ behavior of $\Phi_{t\over2}$. Classically,
since $t/2$ is negative, the leading term is (see \limphin)
\eqn\ptl{\Phi_{t\over2}(x)={1-t\over\pi}
e^{-\half Qt\phi}|\gamma-x|^{-2t}~.}
As for $\Phi_{-\half}$, we expect to find a multiplicative correction
to this in the full quantum theory. We again use the reflection
symmetry \teschone, and the fact that the dual operator $\Phi_{1-{t\over2}}$
behaves at large $\phi$ like
\eqn\dualphittwo{\Phi_{1-{t\over2}}\simeq
-e^{-\half Qt\phi}\delta^2(\gamma-x)~.}
Substituting this into \teschone\ leads to
\eqn\ptltrue{\Phi_{t\over2}(x)\simeq{1-t\over\pi}\RR({t\over2})
e^{-\half Qt\phi}|\gamma-x|^{-2t}~.}
{}For the choice of $\lambda$ in \lamval,
the reflection coefficient is given by \rrhh, hence
\eqn\lamttwo{\RR({t\over2})
={\Gamma(2-{1\over t})\over \Gamma({1\over t})}~.}
Comparing \ptl\ to \ptltrue\ we see that the quantum correction is
again a multiplicative factor given by the appropriate reflection
coefficient.

Multiplying \ptltrue\ by the leading behavior of $\Phi_h(y)$
given by the second line of \expphi\ gives the free field theory result
\eqn\ptm{\eqalign{
\Phi_{t\over2}(x)\Phi_h(y)=&-{1-t\over\pi}\RR({t\over2})|x-y|^{-2t}
e^{Q(h-{t\over2}-1)\phi}\delta^2(\gamma-y)\cr
=&{1-t\over\pi}\RR({t\over2})|x-y|^{-2t}
\Phi_{h-{t\over2}}(y)~.\cr}}
Comparing to \ooe\ we see that
\eqn\cctwo{C_2={1-t\over\pi}\RR({t\over2})~.}
We next compute $C_1$ in order to derive a recursion
relation similar to \ttt.

As mentioned above, the calculation is perturbative in Sine-Liouville
(or in the supersymmetric case in $N=2$ Liouville). Therefore,
we start by briefly reviewing the map (which is also reviewed in \kkk,
but we use slightly different normalizations). The claim is that
the CFT on the cigar $SL(2)_k/U(1)$, which has central charge
\eqn\ccigar{c={3k\over k-2}-1=2+{6\over k-2}~,}
is equivalent to the CFT on $R_\phi\times S^1_x$,
where $R_\phi$ is the theory of a real scalar $\phi$
with a linear dilaton, and $S^1_x$ is the theory of a scalar
$x=x_l+x_r$ on a circle with radius
\eqn\rcyl{R=\sqrt{2k}~.}
The free part of the Lagrangian $\CL=\CL_0+\CL_{\rm int}$ is
\eqn\fpol{\CL_0=\partial x\bar\partial x+
\partial\phi\bar\partial\phi-Q\widehat R\phi~.}
The linear dilaton slope $Q$ and the corresponding central charge
of the cylinder theory are
\eqn\Qcyl{Q^2={2\over k-2}=-{2\over t};\;\;c=2+3Q^2=2+{6\over k-2}~,}
as in \ccigar. There is an interaction term in the Lagrangian given by
\eqn\lint{\CL_{\rm int}
=\lambda_{sl}\cos{R\over 2}(x_l-x_r)e^{\half Qt\phi}~.}
One can check that it has dimension $(1,1)$.

The operator map between the observables on the cigar \primr\
(for simplicity for $m=\bar m$, which is all we need here)
and Sine-Liouville is
\eqn\cigsin{V_{j;m,m}\leftrightarrow e^{ip(x_l-x_r)+\beta\phi}~,}
with
\eqn\pbj{\eqalign{p=&m\sqrt{2\over k}~,\cr
\beta=&Qj~.\cr
}}
One can check that
\eqn\matchd{\half p^2-\half\beta(\beta+Q)={m^2\over k}-{j(j+1)\over k-2}~.}
Note that KPZ scaling implies that
\eqn\vvvjjjss{\langle V_jV_j\rangle\sim\lambda_{sl}^{-{2\over t}(2j+1)}~.}
Comparing to \scspcor\ we see that
\eqn\lamslsl{\lambda_{sl}^{-{2\over t}}\sim\lambda~.}
Below we will determine the precise relation between the couplings:
\eqn\llsl{\pi\lambda{\Gamma(-{1\over t})\over \Gamma(1+{1\over t})}
=\left({\pi\lambda_{sl}\over t}\right)^{-{2\over t}}~.}

\noindent
We would like to use the Sine-Liouville variables to calculate $C_1$ in
\ooe.
Start by sending $x,y\to 0$. Looking back at \ptltrue\ we see that
\eqn\eea{\Phi_{t\over2}(x=0)={1-t\over\pi}\RR({t\over2})e^{-\half t Q\phi}
(\gamma\bar\gamma)^{-t}~.}
Comparing to \primr\ we see that
\eqn\eeb{\Phi_{t\over2}(x=0)={1-t\over\pi}\RR({t\over2})
V_{-{t\over2};-{t\over2},-{t\over2}}~.}
Similarly, from \ppp\ we learn that
\eqn\eec{\Phi_h(y=0)=V_{h-1;-h,-h}~.}
So, in the limit $x,y\to 0$, the $C_1$ term in \ooe\ becomes
\eqn\eed{{1-t\over\pi}\RR({t\over2})V_{-{t\over2};-{t\over2},-{t\over2}}
V_{h-1;-h,-h}=C_1V_{h+{t\over2}-1;-h-{t\over2},-h-{t\over2}}~.}
This equation should be true also in $SL(2)/U(1)$, since the
OPE of the $U(1)$ part is trivial. Thus we can use the map \cigsin,
\pbj, and write it as (replacing $x_l-x_r$ by $x$ for brevity)
\eqn\eee{{1-t\over\pi}
\RR({t\over2})e^{-i{t\over2}\sqrt{2\over k}x-{t\over2}Q\phi}
e^{-ih\sqrt{2\over k}x+(h-1)Q\phi}=
C_1e^{-i(h+{t\over2})\sqrt{2\over k}x+(h+{t\over2}-1)Q\phi}~.}
We see that the l.h.s. and r.h.s. have the same $p$ but there is
a mismatch of $tQ$ in the value of $\beta$ (in the notation of the
r.h.s. of \cigsin). This is easy to fix; we simply expand
$\exp(-\int \CL_{\rm int})$ \lint\ to second order in $\lambda_{sl}$:
\eqn\eef{\eqalign{
&{1-t\over\pi}\RR({t\over2}){\lambda_{sl}^2\over2}\int d^2z_1\int d^2z_2
\cos{\sqrt{k\over2}x(z_1)} e^{\half tQ\phi(z_1)}
\cos{\sqrt{k\over2}x(z_2)} e^{\half tQ\phi(z_2)}\cr
&e^{-i{t\over2}\sqrt{2\over k}x-{t\over2}Q\phi}
e^{-ih\sqrt{2\over k}x+(h-1)Q\phi}=
C_1e^{-i(h+{t\over2})\sqrt{2\over k}x+(h+{t\over2}-1)Q\phi}~.
}}
Performing the free field OPE's as before we find that $C_1$ is given by
(setting $z=1$)
\eqn\eeg{C_1=\pi(1-t)\RR({t\over2})\lambda_{sl}^2\int d^2z_1d^2z_2
|z_1-z_2|^{2(1-k)}|1-z_1|^{2(k-2)}|z_1|^{-2}
|z_2|^{2(2h-1)}~.}
Since the power of $1-z_2$ vanishes, we can perform the integral
over $z_2$, and then that over $z_1$, using standard formulae.
One finds that all the $\Gamma$ functions cancel and
\eqn\eeh{C_1(h)=-{\pi(1-t)\lambda_{sl}^2\over (2h+1-k)^2}\RR({t\over2})=
-{\pi(1-t)\lambda_{sl}^2\over (2h+t-1)^2}\RR({t\over2})~.}
We can now derive a recursion relation for the two point function
$D(h)$ \twoptfn\ by following the same logic as in eqs \thpt\ --
\ttt\ by considering the three point function
$$\langle\Phi_{t\over2}(x)\Phi_h(x_1)\Phi_{h+{t\over2}}(x_2)\rangle~.$$
One finds:
\eqn\eei{{D(h)\over D(h+{t\over2})}={C_1(h)\over C_2(h+{t\over2})}~.}
Plugging in our results \cctwo, \eeh\ we find
\eqn\eej{{D(h)\over D(h+{t\over2})}=-{\pi^2\lambda_{sl}^2\over(2h+t-1)^2}~.}
The correct answer \corans\ satisfies
\eqn\eek{{D(h)\over D(h+{t\over2})}=-{t^2\nu^{-t}\over(2h+t-1)^2}~.}
Clearly, we can set $\lambda_{sl}$ to a value
such that the r.h.s. of \eek\ coincides with that of \eej:
\eqn\setlsl{\nu=\left({\pi\lambda_{sl}\over t}\right)^{-{2\over t}}~.}
Comparing to \nulagg\ we see that the relation between $\lambda_{sl}$
and $\lambda$ is indeed as advertised in eq. \llsl.
In particular, for the choice \lamval\ of the normalization of
the operators, we have $\nu=1$ thus
\eqn\sineliocoup{\lambda_{sl}={t\over\pi}~.}

To summarize, equations \ttt\ and \eej\ determine
the two point function $D(h)$ uniquely, at least in the case
when $k$ (or $t$) is irrational. Of course, the calculations
described here can also be viewed as evidence for the strong-weak
coupling duality of Sine-Liouville and the cigar CFT.

So far we have only determined $C_1$ and $C_2$ in \ooe. $C_3$ can be
calculated by using the reflection symmetry \teschone.
Substituting \teschone\ into \ooe\ we get
\eqn\dddkkk{\eqalign{
&\RR(h){2h-1\over\pi}\Phi_{t\over2}(x;z)
\int d^2y'|y-y'|^{-4h}\Phi_{1-h}(y';w)=\cr
&C_1(h)\Phi_{h+{t\over2}}(y)
+C_2(h)|x-y|^{-2t}\Phi_{h-{t\over2}}(y)+C_3(h)|x-y|^{2(1-t-2h)}
\Phi_{1-{t\over2}-h}(y)~.\cr}}
Using \ooe\ directly on the l.h.s. gives
\eqn\lhsope{\eqalign{
&\RR(h){2h-1\over\pi}\int d^2y'|y-y'|^{-4h}\big[
C_1(1-h)\Phi_{1-h+{t\over2}}(y')+\cr
&C_2(1-h)|x-y'|^{-2t}\Phi_{1-h-{t\over2}}(y')+C_3(1-h)
|x-y'|^{2(2h-t-1)}\Phi_{h-{t\over2}}(y')\big]~.\cr}}
The second term in \lhsope\ has the same behavior as the
third term on the r.h.s. of \dddkkk.~\foot{Similarly,
the third term in \lhsope\ has the same behavior as the
second term on the r.h.s. of \dddkkk. On the contrary,
it is not manifest how to identify the first term
on the r.h.s of \dddkkk\ with the first term in \lhsope.}
By comparing the two one can
compute $C_3$. One finds:
\eqn\cthreeans{C_3(h)={1\over\pi}(2h-1)(1-t)\RR({t\over2})
{\Gamma(1+{2h-1\over t})\Gamma(1-2h)\Gamma(1-t)\Gamma(2h+t-1)
\over\Gamma(1-{2h-1\over t})\Gamma(2h)\Gamma(t)\Gamma(2-2h-t)}~.}
As a check, note that $C_3$ satisfies the constraint coming from
$\langle\Phi_{t\over2}\Phi_h\Phi_{1-h-{t\over2}}\rangle$,
\eqn\cccthree{C_3(h)D(1-h-{t\over2})=C_3(1-h-{t\over2})D(h)~.}

\newsec{Degenerate conformal blocks on the sphere}

In this section we review the calculation of current algebra blocks
relevant for the four point functions of the degenerate operators
$\Phi_{-\half}$ and $\Phi_{t\over2}$ with operators with generic $h$.
This provides a check on the fusion coefficients computed in the
previous section; it is also needed for the study of D-branes in \gksc.

We start by deriving the generalization \ZamolodchikovBD\ of the
Knizhnik-Zamolodchikov (KZ) equation for $SL(2)$ CFT. The four point
function of $\Phi_{h_j}(x_j, \bar x_j;z_j, \bar z_j)$ $(j=1,2,3,4)$
can be written using worldsheet and spacetime conformal invariance
in the following form (see \tesss\ eqs. (44) -- (48)):
\eqn\donea{\eqalign{
&\langle \Phi_{h_1}(x_1,\bar x_1;z_1, \bar z_1)\cdots
\Phi_{h_4}(x_4,\bar x_4;z_4, \bar z_4)\rangle=\cr
&|z_1-z_4|^{2(\Delta_3+\Delta_2-\Delta_1-\Delta_4)}
|z_3-z_4|^{2(\Delta_1+\Delta_2-\Delta_3-\Delta_4)}\cr
&|z_2-z_4|^{-4\Delta_2}|z_1-z_3|^{2(\Delta_4-\Delta_1-\Delta_2-\Delta_3)}\cr
&|x_1-x_4|^{2(h_3+h_2-h_1-h_4)}
|x_3-x_4|^{2(h_1+h_2-h_3-h_4)}\cr
&|x_2-x_4|^{-4h_2}|x_1-x_3|^{2(h_4-h_1-h_2-h_3)}
\CF(\eta_{\rm ws},\eta_{\rm st})~.\cr
}}
Here $\eta_{\rm ws}$ and $\eta_{\rm st}$ are the worldsheet and spacetime
crossratios,
\eqn\dtwo{\eqalign{
\eta_{\rm ws}=&{(z_1-z_2)(z_3-z_4)\over (z_1-z_3)(z_2-z_4)}~,\cr
\eta_{\rm st}=&{(x_1-x_2)(x_3-x_4)\over (x_1-x_3)(x_2-x_4)}~.\cr
}}
Note that if the dimensions are equal in pairs,
\eqn\dthree{\eqalign{
&\Delta_1=\Delta_3;\quad \Delta_2=\Delta_4~,\cr
&h_1=h_3;\quad h_2=h_4~,\cr
}}
\donea\ simplifies:
\eqn\dfour{\eqalign{
&\langle \Phi_{h_1}(x_1,\bar x_1;z_1, \bar z_1)
\Phi_{h_2}(x_2,\bar x_2;z_2, \bar z_2)
\Phi_{h_1}(x_3,\bar x_3;z_3, \bar z_3)
\Phi_{h_2}(x_4,\bar x_4;z_4, \bar z_4)\rangle=\cr
&|z_2-z_4|^{-4\Delta_2}|z_1-z_3|^{-4\Delta_1}
|x_2-x_4|^{-4h_2}|x_1-x_3|^{-4h_1}\CF(\eta_{\rm ws},\eta_{\rm st})~.\cr
}}
We will next compute $\CF$ for the two cases
of interest to us: $h_1=h_3=-1/2$ and $h_1=h_3=t/2$ with $h_2=h_4=h$
in both cases.

\subsec{The current algebra blocks for $h_1=h_3=-1/2$}

Substituting $$h_1=h_3=-\half; \;\;h_2=h_4=h$$ in \dfour, we have:
\eqn\dfive{\eqalign{
&\langle \Phi_{-\half}(x_1,\bar x_1;z_1, \bar z_1)
\Phi_{h}(x_2,\bar x_2;z_2, \bar z_2)
\Phi_{-\half}(x_3,\bar x_3;z_3, \bar z_3)
\Phi_{h}(x_4,\bar x_4;z_4, \bar z_4)\rangle=\cr
&|z_2-z_4|^{-4\Delta_h}|z_1-z_3|^{-4\Delta_{-\half}}
|x_2-x_4|^{-4h}|x_1-x_3|^2\CF(\eta_{\rm ws},\eta_{\rm st})~.\cr
}}
Due to \degeq, $\CF$ must be a sum of terms like
\eqn\dsix{\CF(\eta_{\rm ws},\eta_{\rm st})=
|\CF_0(\eta_{\rm ws})+\eta_{\rm st}\CF_1(\eta_{\rm ws})|^2~.}
The differential equation that we will derive will act on each
factor separately. Therefore, from now on we can focus on the holomorphic
part of $\CF$ in \dsix:
\eqn\dseven{\eqalign{
&\langle\Phi_{-\half}\Phi_h\Phi_{-\half}\Phi_h\rangle=
(z_2-z_4)^{-2\Delta_h}(z_1-z_3)^{-2\Delta_{-\half}}\cr
&\left[(x_2-x_4)^{-2h}(x_1-x_3)\CF_0(\eta_{\rm ws})+
(x_2-x_4)^{-2h-1}(x_1-x_2)(x_3-x_4)\CF_1(\eta_{\rm ws})\right]~.\cr
}}
We would next like to compute $\CF_0$, $\CF_1$ by solving the KZ equation for
the four point function \dseven.

Let us review the derivation of the KZ equation for this case.
The worldsheet stress tensor is (see \ks\ eq. (2.26))
\eqn\deight{T^{\rm ws}=
-{1\over 2t}\left[J\partial_x^2 J-\half(\partial_x J)^2\right]~.}
Consider
\eqn\dnine{\langle T^{\rm ws}(z)\Phi_{h_1}(x_1;z_1)
\cdots\Phi_{h_4}(x_4;z_4)\rangle~,}
and focus on the coefficient of $1/(z-z_1)$. On the one hand we have
\eqn\dten{\langle T^{\rm ws}(z)\Phi_1\cdots\Phi_4\rangle\sim {1\over z-z_1}
\langle\partial_{z_1}\Phi_1\Phi_2\Phi_3\Phi_4\rangle~.}
On the other hand, substituting \deight\ and using the OPE's \JPhi\
we get\foot{Note that $T^{\rm ws}$ is independent of $x$ and hence
we can choose $x=x_1$ when plugging \deight\ in \dten.}
\eqn\deleven{\eqalign{
&\langle T^{\rm ws}(z)\Phi_1\cdots\Phi_4\rangle\sim
-{1/t\over z-z_1}\sum_{i=2}^4{1\over z_1-z_i}\times\cr
&\left\{\left[(x_i-x_1)^2
{\partial\over\partial x_i}+2h_i(x_i-x_1)\right]
\langle\partial_{x_1}\Phi_1\Phi_2\Phi_3\Phi_4\rangle
-2h_1\left[(x_i-x_1){\partial\over\partial x_i}+h_i\right]
\langle\Phi_1\cdots\Phi_4\rangle\right\}\cr
}}
Comparing the two gives the KZ equation \ZamolodchikovBD\
\eqn\dtwelve{
-t{\partial\over\partial z_1}\langle\Phi_1\cdots\Phi_4\rangle=
\sum_{i=2}^4{1\over z_1-z_i}Q_i\langle\Phi_1\cdots\Phi_4\rangle~,}
where
\eqn\dddttt{Q_i=(x_i-x_1)^2{\partial^2\over\partial x_i\partial x_1}+
2(x_i-x_1)(h_i\partial_{x_1}-h_1\partial_{x_i})-2h_1h_i~.}
Substituting \dseven\ into \dtwelve\ we get the following
first order differential equations for $\CF_0$, $\CF_1$:
\eqn\kzeqns{\eqalign{
t(\eta_{\rm ws}-1)\eta_{\rm ws}{\partial\CF_0\over\partial\eta_{\rm ws}}=&
h\CF_0+\eta_{\rm ws}\CF_1~,\cr
t(1-\eta_{\rm ws})\eta_{\rm ws}{\partial\CF_1\over\partial\eta_{\rm ws}}=&
2h\CF_0+(h-1+2\eta_{\rm ws})\CF_1~.\cr}}
These equations have two independent sets of solutions which
are\foot{We normalized $F_0$ such that for small $\eta_{\rm ws}$
it goes like $F_0\simeq \eta_{\rm ws}^c$ where $c$ is different in the
two cases, but the coefficient of the power is one.}:
\eqn\solone{\eqalign{
&\CF_0^{(-)}
=x^{-a}(1-x)^{-a}F(-2a,2b-2a; b-2a;x)\cr
&=x^{-a}(1-x)^{a-b}F(b,-b;b-2a;x)~,\cr
&\CF_1^{(-)}={2a\over b-2a}x^{-a}(1-x)^{-a}F(1-2a,2b-2a;b-2a+1;x)\cr
&={2a\over b-2a}x^{-a}(1-x)^{a-b}F(b,1-b;b-2a+1;x)~,\cr
}}
\eqn\soltwo{\eqalign{
&\CF_0^{(+)}=x^{a-b+1}(1-x)^{a-b}F(2a-2b+1,2a+1; 2a-b+2;x)\cr
&=x^{a-b+1}(1-x)^{-a}F(1-b,1+b; 2a-b+2;x)~,\cr
&\CF_1^{(+)}={b-2a-1\over b}x^{a-b}(1-x)^{-a}
F(1-b,b;2a-b+1;x)\cr
&={b-2a-1\over b}x^{a-b}(1-x)^{a-b}F(2a-2b+1,2a;2a-b+1;x)~,\cr
}}
where
\eqn\defsol{a\equiv {h\over t};\;\;b\equiv{1\over t};\;\;
x\equiv\eta_{\rm ws}~.}
The function $\CF$ that appears in \dfive, \dsix\ is a linear
combination of the two solutions $\CF_\pm$:
\eqn\daaa{\CF(\eta_{\rm ws},\eta_{\rm st})=
A|\CF_0^{(-)}(\eta_{\rm ws})+\eta_{\rm st}\CF_1^{(-)}(\eta_{\rm ws})|^2
+B|\CF_0^{(+)}(\eta_{\rm ws})+\eta_{\rm st}\CF_1^{(+)}(\eta_{\rm ws})|^2~.
}
$A$ and $B$ can be determined by considering the limit $\eta_{\rm ws}
\to 0$. One finds:
\eqn\dbbb{\eqalign{
A=&|C_-(h)|^2D(h-\half)~,\cr
B=&{|C_+(h)|^2\over (2h-1+t)^2}D(h+\half)~.\cr
}}
A non-trivial check on the structure constants that we derived
above ($C_\pm$) comes from the requirement that if we exchange
$2\leftrightarrow 4$ in \dfive, we get a correct equation (crossing
symmetry). Looking at \dtwo, \dfive\ we see that:
\item{(1)} The prefactors in front of $\CF$ in \dfive\ are invariant
under $2\leftrightarrow 4$.
\item{(2)} The transformation of $\eta_{\rm ws}$, $\eta_{\rm st}$ under
$2\leftrightarrow 4$ is: $\eta\to 1-\eta$.

\noindent
Thus, we conclude that crossing symmetry implies that
(denoting $\eta_{\rm ws}$ by $x$ for brevity):
\eqn\ddddd{\eqalign{
&A|\CF_0^{(-)}(x)+\eta_{\rm st}\CF_1^{(-)}(x)|^2
+B|\CF_0^{(+)}(x)+\eta_{\rm st}\CF_1^{(+)}(x)|^2=\cr
&A|\CF_0^{(-)}(1-x)+(1-\eta_{\rm st})\CF_1^{(-)}(1-x)|^2
+B|\CF_0^{(+)}(1-x)+(1-\eta_{\rm st})\CF_1^{(+)}(1-x)|^2~.
}}
We can write the r.h.s. of \ddddd\ as:
\eqn\dfff{
A|\XX_1-\eta_{\rm st}\XX_2|^2
+B|\XX_3-\eta_{\rm st}\XX_4|^2~,}
which defines $\XX_i$, $i=1,2,3,4$.
In order for \ddddd\ to be valid it seems that we need
a matrix relation of the sort:
\eqn\ddffdd{\eqalign{
\XX_1-\eta_{\rm st}\XX_2=
\tilde a\left(\CF_0^{(-)}(x)+\eta_{\rm st}\CF_1^{(-)}(x)\right)+
\tilde b\left(\CF_0^{(+)}(x)+\eta_{\rm st}\CF_1^{(+)}(x)\right)~,\cr
\XX_3-\eta_{\rm st}\XX_4=
\tilde c\left(\CF_0^{(-)}(x)+\eta_{\rm st}\CF_1^{(-)}(x)\right)+
\tilde d\left(\CF_0^{(+)}(x)+\eta_{\rm st}\CF_1^{(+)}(x)\right)~.\cr
}}
After some algebra we get:
\eqn\aabbccdd{\eqalign{
\tilde a=&{\Gamma(b-2a+1)\Gamma(2a-b)\over\Gamma(b+1)\Gamma(-b)}~,\cr
\tilde b=&{b\over b-2a-1}{\Gamma^2(b-2a)\over\Gamma(-2a)\Gamma(2b-2a)}~,\cr
\tilde c=&{b-2a\over b}
{\Gamma(2a-b+2)\Gamma(2a-b)\over\Gamma(2a+1)\Gamma(2a-2b+1)}~,\cr
\tilde d=&-\tilde a~.\cr
}}
A useful thing to note is that the determinant of the $2\times 2$
matrix we found is $-1$:
\eqn\detabcd{\tilde a\tilde d-\tilde b\tilde c=-1~.}
The fact that this determinant has absolute value one is a necessary
consistency condition. Substituting \aabbccdd\ into the previous
equations we find the constraint
\eqn\cchhdd{
{|C_+(h)|^2 D(h+\half)\over |C_-(h)|^2 D(h-\half)}
={\Gamma^2(b-2a)\Gamma(2a+1)\Gamma(2a-2b+1)\over
\Gamma^2(2a-b+1)\Gamma(-2a)\Gamma(2b-2a)}~.}
Plugging in the explicit formulae we found before for the
l.h.s. we indeed find the r.h.s., which verifies the consistency
of the procedure.

\subsec{The current algebra blocks for $h_1=h_3=t/2$}

In this case we would like to substitute
$$h_1=h_3=t/2;\qquad h_2=h_4=h$$
in \dfour\ and compute the conformal blocks $\CF$.
The degeneracy equation following from the vanishing of $\theta$
\abc\ takes in this case the form
\eqn\degeqnt{(\DD_2+z\DD_3)\CF=0~,}
with
\eqn\ddtwothree{\eqalign{
\DD_2=&(e^y-1)(t+\partial_y)(2h+\partial_y)(t+1+\partial_y)+
2(t+\partial_y)[h(t+1)+\partial_y]~,\cr
\DD_3=&(e^{-y}-1)\partial_y^2(\partial_y-1)+t(t-1)\partial_y-
2\partial_y^2~,\cr
}}
where $x=\exp(y)$.
The KZ equation \dddttt\ takes the form
\eqn\kzttwo{-tz\partial_z\CF=Q_2\CF-{z\over 1-z}Q_4\CF~,}
where
\eqn\qtwofour{\eqalign{
Q_2=&e^y(t+\partial_y)(2h+\partial_y)-
(\partial_y^2-\partial_y+t\partial_y+2h\partial_y+ht)~,\cr
Q_4=&-e^y(2h+\partial_y)(t+\partial_y)+(h+\partial_y)(t+2\partial_y)
-e^{-y}\partial_y^2~.\cr
}}
These equations have three solutions with the boundary conditions that
we are interested in:
\eqn\fabc{\eqalign{
F_A(x;z)=&z^h(1-z)^hF_1(2h,t,2h+t-1;2h+t;x,z)~,\cr
F_B(x;z)=&x^{-t}z^{1-h}(1-z)^hF_1(t,t,1-t;2-2h;{z\over x},z)~,\cr
F_C(x;z)=&z^h(1-z)^he^{-i\pi(1-t)}{\Gamma^2(2h)\over\Gamma(2h+1-t)
\Gamma(2h+t-1)}\times\cr
&\left[Z_8-{\Gamma(2h+1-t)\Gamma(1-2h-t)
\over\Gamma^2(1-t)}e^{2i\pi h}Z_1\right]~,\cr
}}
where
\eqn\zeightone{\eqalign{
Z_8=&x^{-2h}F_1(2h,1-t,t+2h-1;2h+1-t;{1\over x}, {z\over x})~,\cr
Z_1=&F_1(2h,t,t+2h-1;t+2h;x,z)~.\cr
}}
In the last equations we are using the hypergeometric function
in two variables (whose definition and some properties are summarized
in the appendix).

As $x,z\to 0$, the conformal blocks \fabc\ have the following
asymptotic behavior\foot{More precisely, we are discussing here
the limit $z\to 0$ first, followed by $x\to 0$. The order of limits
is important in this case.}:
\eqn\asbeh{\eqalign{
F_A&\simeq z^h~,\cr
F_B&\simeq x^{-t}z^{1-h}~,\cr
F_C&\simeq x^{1-t-2h}z^h~.\cr
}}
The solution for the four point function on the sphere \dfour\
has the form
\eqn\cccfff{\CF(x;z)=A|F_A(x;z)|^2+B|F_B(x;z)|^2+C|F_C(x;z)|^2~,}
where the coefficients $A$, $B$, $C$ are obtained by sending
$x,z\to 0$ (in the order indicated in footnote 10) and comparing
the behavior to the contributions of the terms in \ooe. This leads
to:
\eqn\AABBCC{\eqalign{
A=&|C_1(h)|^2D(h+{t\over2})~,\cr
B=&|C_2(h)|^2D(h-{t\over2})~,\cr
C=&|C_3(h)|^2D(1-h-{t\over2})~.\cr
}}
Again, a non-trivial check on the structure constants $C_{1,2,3}$
can be done by requiring crossing symmetry:
\eqn\crosym{|\CF(1-x;1-z)|=|\CF(x;z)|~,}
for the blocks in eq. \cccfff.
Defining a $3\times 3$ matrix $\MM$ by:
\eqn\tedious{F_I(1-x;1-z)=\MM_{IJ}F^J(x;z)~, \qquad I,J=A,B,C~,}
one finds after some algebra that
\eqn\MAABBCC{\eqalign{
\MM_{AA}=&{\sin\pi t\over\sin\pi(2h+t)}~,\cr
\MM_{AB}=&{1-2h-t\over 1-2h}~,\cr
\MM_{AC}=&{\Gamma(2h+t)\Gamma(2h+t-1)\Gamma(1-2h)\over
          \Gamma^2(t)\Gamma(2h)}~,\cr
\MM_{BA}=&{1-2h\over 1-2h-t}~,\cr
\MM_{BB}=&-e^{i\pi t}{\sin\pi t\over\sin 2\pi h}~,\cr
\MM_{BC}=&-e^{i\pi(2h+t)}{\Gamma(1-2h)\Gamma(2-2h)\Gamma(2h+t-1)\over
          \Gamma^2(t)\Gamma(2-2h-t)}~,\cr
\MM_{CA}=&{\Gamma(2-2h-t)\Gamma(1-2h-t)\Gamma(2h)\over
          \Gamma^2(1-t)\Gamma(1-2h)}~,\cr
\MM_{CB}=&-e^{i\pi(2h+t)}{\Gamma(2h)\Gamma(2h-1)\Gamma(2-2h-t)\over
          \Gamma^2(1-t)\Gamma(2h+t-1)}~,\cr
\MM_{CC}=&e^{2\pi i(h+t)}+e^{i\pi(2h+t)}
         {\sin^2\pi t\over\sin\pi(2h+t)\sin 2\pi h}~.\cr
}}
These $\MM_{IJ}$ are the same as the ones found by J. Teschner; they
can be read off from subsection 7.3.1 in \tesch.
Using this $\MM$ in \cccfff, \crosym, and comparing the coefficients
$A,B,C$ obtained this way to eq. \AABBCC, one finds the same absolute
values of $C_{1,2,3}$ as derived above.

\newsec{$N=2$ superconformal extension}

In the supersymmetric case, the duality of \fzzsl\ is replaced by
the conjectured equivalence of $N=2$ Liouville and the supersymmetric
cigar CFT \gk. In this section we discuss calculations analogous to those
performed in section 4, for the supersymmetric system.
Since the calculations are similar to those of section 4, we will be
rather schematic and focus mainly on the differences.

In the superconformal coset model $SL(2)/U(1)$ there are observables
$V_{j;m,\bar m}$ with scaling dimensions
\eqn\hofv{\Delta(V_{j;m,\bar m})={j(j+1)-m^2\over t}~,}
where, as before,
\eqn\tkmt{t\equiv -(k-2)~.}
The right moving scaling dimension of $V_{j;m,\bar m}$ is given by
a formula similar to \hofv, with $m\to\bar m$. $k$ is the bosonic level
of the $SL(2)/U(1)$ sigma model while $k-2=-t$ is the total level.

Far from the tip of the cigar this SCFT looks like a
sigma model on the cylinder $R_\phi\times S^1_Y$, where
$R_\phi$ is the real line with a linear dilaton
\eqn\ldil{\Phi(\phi)=-{Q\over 2}\phi~,}
with $Q$ given in \Qk, and the circle $S^1_Y$ is parametrized by a
canonically normalized scalar $Y$ with radius
\eqn\rrcyl{R_Y=\sqrt{-2t}~.}
The observables $V_{j;m,\bar m}$ \hofv\ have the asymptotic form
\eqn\asfv{V_{j;m,\bar m}\to e^{Q(j\phi+imY+i\bar m\bar Y)}~.}
The two point functions of these observables can be computed as in the
bosonic case (see \refs{\gk,\gktwo}).
One starts with correlators in the underlying $SL(2)$
SCFT; those are not affected by adding free fermions
to the bosonic sigma model.
Hence also the two point functions in the $SL(2)/U(1)$ quotient
are not changed relative to the bosonic case,
giving rise again to the relation \ttt.

On the other hand, the $N=2$ Liouville interaction (the top component of
the superpotential plus its complex conjugate) is
\eqn\ntwol{V_L=\mu \Psi\bar\Psi e^{-{1\over Q}(\phi+iY)}+c.c.~,}
where
\eqn\fer{\Psi=\psi^\phi+i\psi^Y~,}
and the fermions $\psi^Y, \psi^\phi$
are the superpartners of the scalars $Y, \phi$, respectively.
Following the same steps as in the previous sections, with the slight changes
discussed above, leads to (the supersymmetric analog of \eef):
\eqn\aeef{\eqalign{
&\RR({t\over 2}){(1-t)|\mu|^2\over2\pi}\prod_{i=1}^2\int d^2z_i
\left(\Psi\bar\Psi(z_1) e^{-{1\over Q}(\phi+iY)(z_1)}+c.c.\right)
\left(\Psi\bar\Psi(z_2) e^{-{1\over Q}(\phi+iY)(z_2)}+c.c.\right)\cr
&e^{-{t\over2}Q(\phi+iY)}e^{(h-1)Q\phi-ihQY}=
C_1e^{(h+{t\over2}-1)Q\phi-i(h+{t\over2})QY}~.
}}
Performing the free field OPE's we find the same $C_1$ as in eq. \eeg,
leading to \eeh\ and \eej.

\bigskip
\noindent{\bf Acknowledgements:}
We thank A. Schwimmer for collaboration on this project. We also thank
J. Maldacena, A. Parnachev and D. Sahakyan for discussions. The work of A.G.
is supported in part by
BSF -- American-Israel Bi-National Science Foundation,
the Israel Academy of Sciences and Humanities --
Centers of Excellence Program,
the German-Israel Bi-National Science Foundation,
and the European RTN network HPRN-CT-2000-00122.
The work of D.K. is supported in part by DOE grant \#DE-FG02-90ER40560.
A.G. thanks the Einstein Center at the Weizmann Institute for partial
support. D.K. thanks the Weizmann Institute for hospitality during
the course of this work.

\appendix{A}{Some useful formulae}

A useful integral is:
\eqn\usefint{\eqalign{\int d^2x|x|^{2a}x^n|1-x|^{2b}(1-x)^m
&=\pi{\Gamma(a+n+1)\Gamma(b+m+1)\Gamma(-a-b-1)\over
\Gamma(-a)\Gamma(-b)\Gamma(a+b+m+n+2)}~,\cr
n,m&\in Z~,}}
and the Gamma functions satisfy:
\eqn\gsat{\Gamma(a+1)=a\Gamma(a)~.}
The hypergeometric function is defined by the differential
equation for a function $u(x)$
\eqn\AAA{x(1-x)u''+\left[\gamma-(\alpha+\beta+1)x\right]u'-\alpha
\beta u=0~.}
This equation has two solutions:
\eqn\AAB{\eqalign{
u_1=&F(\alpha,\beta;\gamma;x)~,\cr
u_2=&x^{1-\gamma}F(\alpha-\gamma+1,\beta-\gamma+1;2-\gamma;x)~.\cr
}}
For small $x$, $F$ can be expanded as follows:
\eqn\AAC{F(\alpha,\beta;\gamma;x)=
\sum_{n=0}^{\infty}{(\alpha)_n(\beta)_n\over n!(\gamma)_n}x^n
=1+{\alpha\beta\over\gamma}x+
{\alpha(\alpha+1)\beta(\beta+1)\over2\gamma(\gamma+1)}x^2+\cdots~,}
where
\eqn\AACA{(a)_n\equiv a(a+1)\cdots (a+n-1)={\Gamma(a+n)\over\Gamma(a)}~.}
Two other identities that are sometimes useful are:
\eqn\AAD{\eqalign{
&F(\alpha,\beta;\gamma;x)=(1-x)^{\gamma-\alpha-\beta}
F(\gamma-\alpha,\gamma-\beta; \gamma;x)~,\cr
&{\partial F\over\partial x}(\alpha,\beta;\gamma;x)={\alpha\beta\over\gamma}
F(\alpha+1,\beta+1;\gamma+1;x)~.\cr
}}
Under $x\to 1/x$:
\eqn\AAE{\eqalign{
&F(\alpha,\beta;\gamma;x)=
{\Gamma(\gamma)\Gamma(\beta-\alpha)\over\Gamma(\beta)\Gamma(\gamma-\alpha)}
(-{1\over x})^\alpha F(\alpha,\alpha+1-\gamma;\alpha+1-\beta;{1\over x})
+\cr
&{\Gamma(\gamma)\Gamma(\alpha-\beta)\over\Gamma(\alpha)\Gamma(\gamma-\beta)}
(-{1\over x})^\beta F(\beta,\beta+1-\gamma;\beta+1-\alpha;{1\over x})~.
\cr}}
Under $x\to 1-x$:
\eqn\AAF{\eqalign{
&F(\alpha,\beta;\gamma;1-x)=
{\Gamma(\gamma)\Gamma(\gamma-\beta-\alpha)\over
\Gamma(\gamma-\alpha)\Gamma(\gamma-\beta)}
F(\alpha,\beta;\alpha+\beta+1-\gamma;x)
+\cr
&x^{\gamma-\alpha-\beta}
{\Gamma(\gamma)\Gamma(\alpha+\beta-\gamma)\over
\Gamma(\alpha)\Gamma(\beta)}
F(\gamma-\alpha,\gamma-\beta;\gamma+1-\alpha-\beta;x)~.
\cr}}
The hypergeometric function in two variables $F_1(x,y)$ can be defined
as the analytic continuation of the small $x,y$ ($|x|,|y|<1$)
expansion:
\eqn\foneex{F_1(\alpha,\beta,\beta';\gamma;x,y)=
\sum_{m=0}^{\infty}\sum_{n=0}^{\infty}
{(\alpha)_{m+n}(\beta)_m(\beta')_n\over m!n!(\gamma)_{m+n}}x^m y^n~,}
where $(a)_n$ is defined in \AACA.

More useful identities:
\eqn\AAG{\Gamma(x)\Gamma(1-x)={\pi\over\sin(\pi x)}~,}
\eqn\AAI{\sin x + \sin y = 2\sin{x+y\over 2}\cos{x-y\over 2}~.}

\listrefs
\end